\documentclass[amsmath,amssymb,aps,pre,twocolumn]{revtex4-2}
\usepackage[utf8]{inputenc}
\usepackage[T1]{fontenc}
\usepackage{amsmath}
\usepackage{graphicx}
\usepackage{booktabs}
\usepackage{svg}
\usepackage{natbib}
\usepackage{multibib}
\usepackage{bbm}
\usepackage{adjustbox}
\usepackage{algorithm}
\usepackage{algpseudocode}
\usepackage{hhline}
\graphicspath{ {Images/} }
\usepackage{array,multirow}
\newcolumntype{P}[1]{>{\centering\arraybackslash}p{#1}}
\renewcommand{\dbltopfraction}{0.9}
\renewcommand{\floatpagefraction}{.8}%

\newcites{SM}{Supplemental References}

\begin{document}
	
	\title{Hierarchical route to the emergence of leader nodes in real-world networks}
	\author{Joseph D.~\surname{O'Brien}}
	\affiliation{MACSI, Department of Mathematics and Statistics, University of Limerick, Limerick V94 T9PX, Ireland}
	\author{Kleber A. Oliveira}
	\affiliation{MACSI, Department of Mathematics and Statistics, University of Limerick, Limerick V94 T9PX, Ireland}
	\author{James P. Gleeson}
	\affiliation{MACSI, Department of Mathematics and Statistics, University of Limerick, Limerick V94 T9PX, Ireland}
	\author{Malbor Asllani}
	\affiliation{MACSI, Department of Mathematics and Statistics, University of Limerick, Limerick V94 T9PX, Ireland}
	
	\begin{abstract}
	 A large number of complex systems, naturally emerging in various domains, are well described by directed networks, resulting in numerous interesting features that are absent from their undirected counterparts. Among these properties is a strong non-normality, inherited by a strong asymmetry that characterizes such systems and guides their underlying hierarchy. In this work, we consider an extensive collection of empirical networks and analyze their structural properties using {information-theoretic} tools. A ubiquitous feature {is observed} amongst such systems as the level of non-normality increases. {When the non-normality reaches a given threshold, highly directed substructures {aiming} {towards} terminal {(sink or source)} nodes, denoted here as \textit{leaders}, spontaneously emerge. Furthermore, the relative number of leader nodes describe the level of anarchy that characterizes the networked systems.} {Based on the structural analysis, we develop} a null model to capture features such as the aforementioned transition {in the networks' ensemble}. 
	 {We also demonstrate} that the role {of leader nodes} {at} the pinnacle of the hierarchy is crucial in driving dynamical processes in these systems. This work paves {the} way for a {deeper} understanding of the architecture of empirical {complex} systems {and the processes taking place on them.}
	\end{abstract}
	\maketitle
	
	\noindent
	Many  real systems in nature are organized such that they are constituted by single entities that interact with one another through complex structures. The architecture of these {interactions} {has} been the subject of study within the field of network science over the past two decades~\cite{newman2010networks, boccaletti2006complex}. 
	{Recently,}  the directedness and hierarchical nature of real networks {has attracted significant focus}~\cite{johnson2017looplessness, Johnson2020digraph, kawakatsu2020emergence}. One prime example is {the} literature demonstrating how many real networks, from biological to social, possess both a strong asymmetry and non-normality~\cite{asllani2018structure, Asllani2018PRE, Johnson2020digraph}.
	This ubiquitous property of real networks has resulted in the concept of non-normal networks, described as those whose adjacency matrix $\mathbf{A}$, 
	{is} (strongly) non-normal, by definition implying that $\mathbf{A}^T\mathbf{A} \neq \mathbf{A}\mathbf{A}^T$~\cite{Trefethen2020}. One striking feature of this finding is the implication that empirical networks are structurally similar to directed acyclic graphs (DAG), from which they also inherit their strong non-normality~\cite{asllani2018structure}.
	Previous {results have} {illustrated} the {effects of the non-normality on the collective} dynamics {from} a variety of processes in areas {such as} the {ecosystems stability~\cite{neubert_caswell_1997}}, synchronization of networked electrical devices~\cite{Ravoori2011}, neuronal dynamics~\cite{Hennequin2012}, network resilience~\cite{Asllani2018PRE}, {trophic relationships~\cite{pilgrim2020organisational}}, pattern formation~\cite{Muolo2019, NN_stoch}, {and information transmission~\cite{baggio2020efficient}.} The asymmetric nature of the networks{, a ubiquity within the field of complex systems~\cite{asllani2018structure},} has been shown to result in qualitatively different behavior {for the processes taking place on them} {in comparison to} those observed upon their symmetric counterparts. {In particular, perturbations of a stable state cause a transient growth {proportional to} the level of non-normality in the linear regime, which may result in a permanent instability in {the nonlinear regime}~\cite{Trefethen2020,Asllani2018PRE, asllani2018structure}}.
	
	Motivated by the structural and dynamical properties of non-normal networks, in this paper, we aim to 
	{gain a higher-level understanding of the }
	relationship 
	between {hierarchy and directedness of} networked structures and the behavior of the dynamical processes therein. With this aim we consider a classical measure that bridges these two aspects---the \textit{entropy rate} (ER)~\cite{cover2005information}. {This measure} provides {an estimation of} the inherent randomness{,} {arising from} the underlying network structure, {within a} given {dynamical} process~\cite{gomez2008entropy, ER_crowd}. {The rationale behind the choice of an entropic measure is based {upon} the intuition that real networks with a strongly non-normal {structure} should have {a} lower level of entropy rate {due to} the hierarchical (directed) topology {in comparison to} a general random network~\cite{asllani2018structure}. {In this sense, we view the network as transporting a quantity of mass (e.g., molecules, information, energy,) across it, driven by} {its DAG-like
	structure}. {Through this interpretation}, the non-normality becomes a meaningful measure to quantify the {level of} polarization {within} the {flow's} equilibrium states as an immediate consequence of the {network's} topological properties.
	
	\begin{figure*}[t!]
		\centering
		\includegraphics[width=0.85\textwidth]{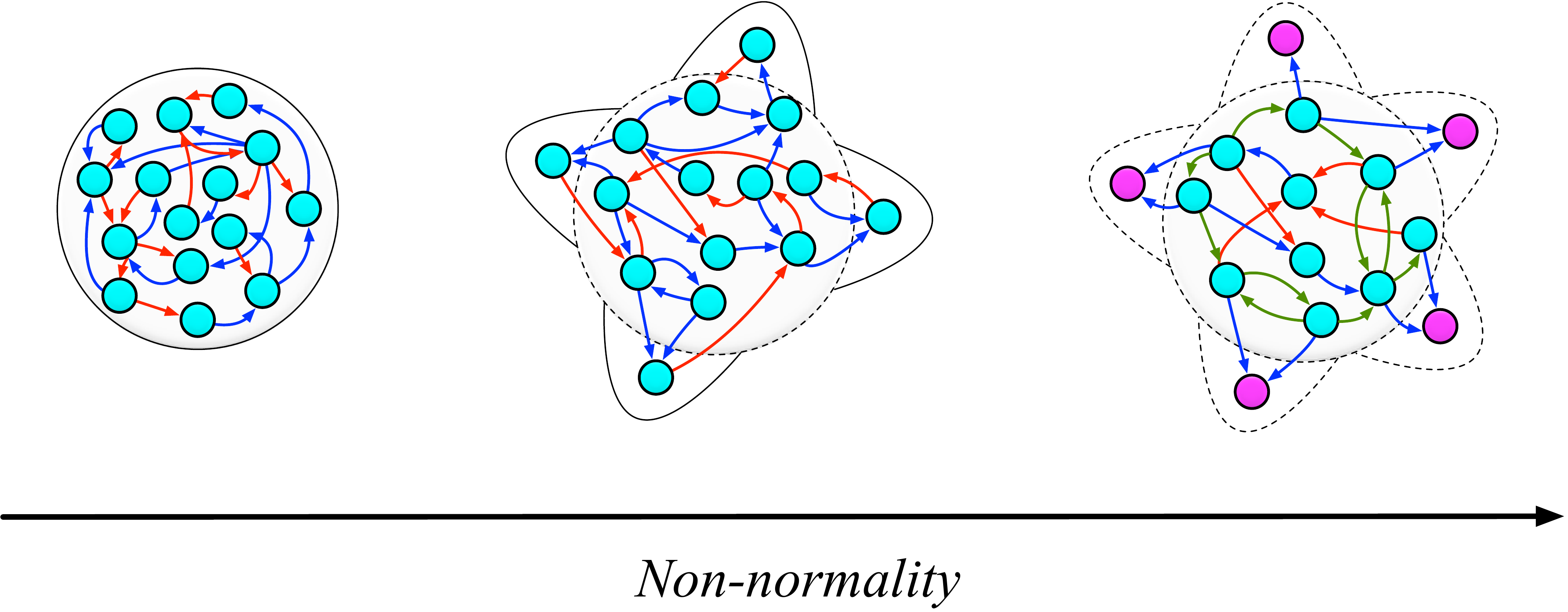}
		\caption{\textbf{The universal emergence of leaders with increasing {non-normality}.} Schematic illustration of the evolution of hierarchical structure within networks ranging from random graphs (left panel) with small non-normality before a hierarchical structure develops in conjunction with the network becoming increasingly {non-normal} (middle panel). And finally when the 
		non-normality surpasses an empirical threshold leader nodes, those with no out-degree at the top hierarchical level (the magenta nodes), {simultaneously emerge across the ensembles of networks}. Blue edges represent links directed up the hierarchical level while red edges are those which move downwards. Also shown are the green edges between nodes at the same hierarchical level.}
		\label{fig:universal_schematic}
	\end{figure*}
	
	With these {ideas} in mind, we proceed to study the entropy rate of the generic random walk taking place on a {large collection of real-world networks from a wide variety of domains. 
	{Surprisingly} we notice that {in the ensemble of networks} the ER {collapses to zero as the non-normality increases}. {This change in ER} is generally monotonic for lower values of non-normality, but once some threshold for higher values of non-normality is {surpassed} the ER {abruptly decreases to zero,} indicating {a sharp transition}. 
	{Such} a pervasive vanishing of the {ER} {measure} within {the ensemble of} empirical networks {is a remarkable occurrence and {is, to our knowledge, a novel} observation.}} 
	To understand this {evidently} universal {phenomenon,} {we first observe that the ER becomes zero when each of the {Strongly Connected Components (SCC)} {in which} mass is trapped {are} constituted by a single node with only incoming edges (sink node) {which in this {paper} we denote as a}} \emph{leader}. 
	{ It is quite} surprising, {firstly,} {that each of} the trapping states {are} simultaneously constituted by sink nodes {(no SCC with multiple nodes co-exist together with leader ones),} and {secondly,} how {this} property depends upon a global {structural} {feature} such as non-normality. A visual illustration of this {behavior} is shown in Fig.~\ref{fig:universal_schematic}. 
	To shed light on the mechanism responsible for the emergence of leader nodes, we extend our analysis to {each network as a whole, looking for}} hierarchical structures. {Once {a} hierarchical ranking of the nodes is obtained} based upon their proximity to the leaders, {an immediate conclusion is reached: real networks share the same {underlying} pattern of hierarchy. In fact, the edges can be classified in three groups: the first, constitute the hierarchical backbone, the DAG substructure made by links that are part of (at least) a directed path to the leaders; the remaining links, {considerably} lower in numbers, show either a common distribution linking higher hierarchical levels with lower ones or create rich-club communities between nodes of the same hierarchy.}
	
{Based on these empirical observations, we propose a mechanistic null model for generating non-normal networks with emerging leader nodes.} 
{Following the recipe proposed in Ref.~\cite{asllani2018structure}}, we {deploy} the classical {Price's model~\cite{price1965networks} {to generate the DAG-backbone links.} Subsequently we complete the model by adding reciprocal links in such a way as to compensate the amount of non-normality, controlled by an external parameter. Our model {accurately captures} the {empirically observed} relation between the entropy rate and {level of} non-normality {within} the {ensemble of networks} and, {importantly}, the abrupt {emergence} of leader nodes.} 
	
{We believe} that {the} ubiquitous emergence of leader nodes is not casual, but a result of an evolutionary process driven by {strong} benefits in the collective dynamics of {interacting individuals}. To emphasize {this idea,} we consider a classical {framework for} {competition} dynamics~\cite{Allee1932,Asllani2018PRE} 
{, and} show that the hierarchical structure results {not only in an obvious benefit for} the leaders {who have considerably higher survivability but also the individuals directly related to {them}.
		
	\section{Results}
	\noindent
    In this study we {aim to understand the relationship that exists between emerging structural properties of real-world networks {with} their} {level of} non-normality {and how such features} affect the resulting dynamical processes which take place upon them. To conduct this analysis we must first provide some quantification of a {network's} non-normality which, in general, implies that the underlying adjacency matrix is such that $\mathbf{A}^T\mathbf{A} \ne \mathbf{A}\mathbf{A}^T$~\cite{Trefethen2020}. {With this aim}, we consider the normalized Henrici departure from normality~{$\hat{d}_F(\mathbf{A}) = \sqrt{||\mathbf{A}||^2_F - \sum_{i = 1}^{N} |\lambda_i|^2}/||\mathbf{A}||_F$, where $||\cdot||_F$ describes the Frobenius norm and $\lambda_i$ represents the eigenvalues of the matrix $\mathbf{A}$~\cite{Trefethen2020}. This quantity varies from the extreme values {of} a symmetric network ($\hat{d}_F(\mathbf{A}) = 0$) to the case of an exact DAG ($\hat{d}_F(\mathbf{A}) = 1$).}
	
	{Since in our analysis the entropy rate will be a diagnostic tool for the level of the {structural} non-normality and the consequent emergence of leaders, we need to formalize its definition with a specific dynamical process. Without loss of generality, we consider in the sequel the generic random-walk that describes the flow of some quantity, which we refer to as mass, that moves between the nodes of a network following the rules specified in the dynamics of the process. We define here the fraction of mass present on node $j$ at time $t$ {to be} given by $q_j(t)$ which may move to the neighbor nodes with some probability dependent on the number of connections the node has~\cite{masuda2017random} (see Methods {for more details}). Specifically, the transition rate~$T_{ij}$ describes the probability {of the particle (constituting the mass, information, etc.)} {moving} from node~$j$ to node~$i$ in each time step. We {describe} this process {via its} stationary distribution $q_j^* = \lim_{t\to\infty} q_j(t)$ which describes the steady state of mass on each node. 
	{The} entropy rate of the process {is successively} given by the following	
	\begin{equation}
		h = - \sum_{i,j} T_{ij} \times q_j^\star \ln(T_{ij}).
		\label{eq:ER}
	\end{equation}	
	{This} quantity provides a gauge regarding the level of randomness of the network structure {through the corresponding transition matrix} associated with a given stochastic dynamical process~\cite{cover2005information,gomez2008entropy}. 
	
	Equipped with these tools 
	we now proceed to consider a large variety of empirical networks and the effect that their structure has upon the corresponding process. Subsequently, we will use the entropy rate as the main observable of a null model for generating synthetic networks that mimic the 
	properties {observed in} the empirical ones.} 
	
	\subsection{Real-world networks}\label{sec:empirical}
	\noindent
{We now consider a} large collection of $124$ {(directed)} real-world networks from a {wide} range of domains {spanning} from biology to social interactions including communication, ecological, transport, {among} many others. To make our analysis compatible with all the networks under scrutiny, we take the following steps. Firstly, since the calculation of the entropy rate requires knowledge of the stationary distribution {which,} in the case of directed graphs, is not necessarily unique~\cite{masuda2017random}, 
we initialize the system {uniformly} such that each node has mass with magnitude given by the reciprocal of network size before proceeding to observe the dynamics until convergence. The second {step we take} is {the} {rescaling of} the entropy rate {in order to make} two distinct networks {comparable with} {one another.} As such, we propose using the following quantity	
	\begin{equation}
	\hat{h} = \frac{h_{\mathbf{A}}}{h_{\text{H}(\mathbf{A})}},
	\end{equation}
	which we coin as the {relative} entropy rate. {The term} $h_{\mathbf{A}}$ describes the entropy rate of the network and $h_{\text{H($\mathbf{A}$)}}$ that of the Hermitian matrix of its adjacency $H(\mathbf{A})~=~(\mathbf{A}~+~\mathbf{A}^T)/2$, which may be viewed as a symmetrized version of the network. {This choice is motivated by the fact that the random walk diffusion tends to accumulate the mass in the nodes
	{of} higher degree.  {Such a rescaling} is crucial in distinguishing the effect of a network structure's directedness upon the equilibrium state of the random walk diffusion in comparison to a related symmetric network}~\cite{newman2010networks, johnson2017looplessness, Johnson2020digraph}. An important property of this measure is related to how we choose the adjacency matrix. In fact, in the {data describing} real networks, {the direction of edges can vary based upon interpretation}, namely both the adjacency matrix $\mathbf{A}$ and its transpose $\mathbf{A}^T$ may be eligible, according to what an outgoing (incoming) edge physically means. To avoid this eventual mismatching and uniformize our measure for all the networks, we choose the direction of edges that minimize $\hat{h}$.
	{Lastly, we highlight that following this agreement a sink node can behave as a source node and vice versa depending upon the interpretation of directionality.}

	\begin{figure*}[t!]
		\centering
		\includegraphics[width=0.8\textwidth]{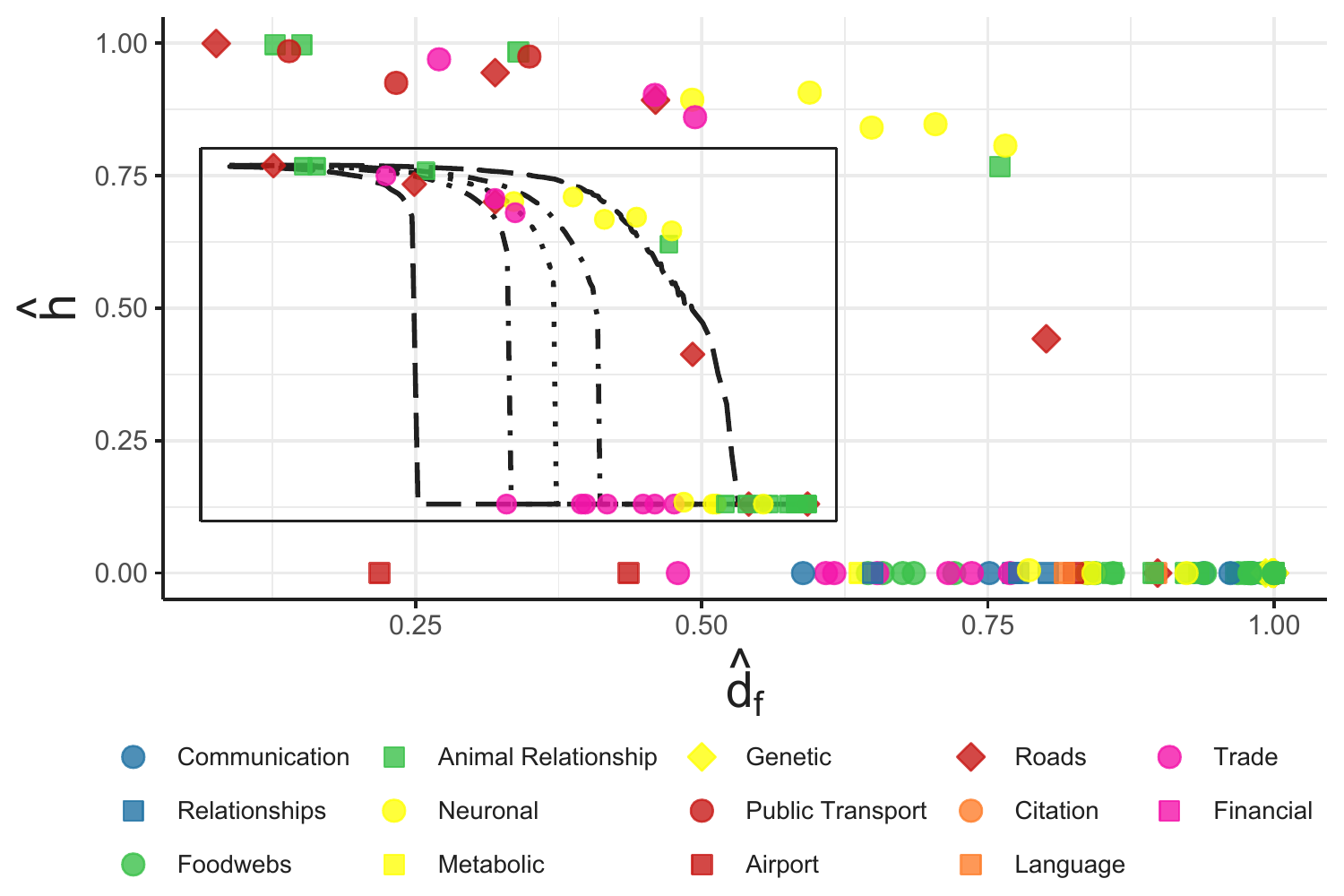}
		\caption{\textbf{Normalized entropy rate {dependence on the network polarization in} real-world networks.} The {network non-normality, quantified by the }normalized Henrici departure $\hat{d}_f$ versus the normalized entropy rate $\hat{h}$ for $124$ empirical networks from a {large} range of domains is shown. {At first glance, the set of networks seem to be grouped in a consistent majority ($\sim85\%$) {having} $\hat{h}=0$, and a small minority ($\sim15\%$) with $\hat{h}\neq 0$. }In particular, {for {those data with} non-zero entropy rate,} a {monotonically decreasing relationship of} the entropy rate with the non-normality is observed. {When some threshold value of the Henrici measure is reached {across the ensemble of networks},} a
		{transition-like} behavior occurs for four of the subdomains {under consideration}, which results in {an abrupt} collapse of the entropy rate towards zero. {{This occurrence} indicates that the mass is accumulated exclusively in single nodes without outgoing edges, {or} leaders. \textbf{Inset:} We have shown the four subdomains where the transition occurs (neuronal, trade, animal relationship, and roads) for a more detailed inspection of the emergent behavior.} Also shown are a number of percentiles {(0.5\%, 25\%, 50\%, 75\%, and 99.5\%)} of {the two quantities} obtained from $10^4$ realizations of the {proposed} {null model} {(N = 100)} with increasing threshold. These realizations indicate  how {our} model {manifests} a similar {transition} in the entropy rate with increasing non-normality.}
		\label{fig:entropy_rate_emp}
	\end{figure*}

	{We present} in Fig.~\ref{fig:entropy_rate_emp} the results of this simulation {for the entire dataset of empirical networks we have collected}. {The first fact that can immediately be noticed is that for most of the networks (more precisely $85\%$ of them), the entropy rate {equals zero}. 
	Such a result is {surprising} since a value of entropy rate that equals zero {implies} that the mass has been accumulated {predominantly} in sink nodes (nodes with only incoming edges)~\footnote{{An exemption occurs when each node of the terminal SCC has a single outgoing edge, namely forms a closed directed path. Network motifs such as closed directed paths are very rare in real-world networks~\cite{Milo_2002} mainly due to the inherent directedness of real networks~\cite{Dominguez_2014}}}. {In the analysis to follow, we shall call these nodes leaders}. The reader can {readily} verify this by referring to Eq.~\eqref{eq:ER} and the Methods section. The other interesting fact, is that for the remaining $15\%$ of the {networks} {there}} appears to exist a negative correlation between the {non-normality} and {the} corresponding normalized entropy rate. In fact, the most remarkable {{finding} is that} there appears to be an {{orderly} monotonic decrease of the ER for four families of networks, {namely those describing}  roads, trade, neuronal, and animal relationships, {in which there is a} collapse of entropy values once the {level of }non-normality {underlying} the networks within said domains surpasses a certain value. An exception of this trend is observed only in the subdomain describing levels of travel between airports, but we {view} this outcome as abnormal due to the physical structure of the networks \footnote{Although they have a relatively small non-normality due to generally having reciprocal (albeit weighted) links, there are interestingly some airports from which passengers only arrive (sink nodes), {resulting in an} anomalous {decrease} in the entropy rate. We believe that this is because such systems are open (the passengers leave from other means {in  the case of} sink airports), making them unsuitable for consistent analysis.}. {These two important aspects that} characterize the empirical networks in our dataset} raise the question, which we proceed to consider {in the section to follow, {regarding what are} the underlying mechanisms {resulting in} the simultaneous
	emergence of leaders in real-world networks and why their occurrence is so ubiquitous {among empirical systems}. 
	
	\begin{figure*}[t]
		\centering		
		\includegraphics[width= \textwidth]{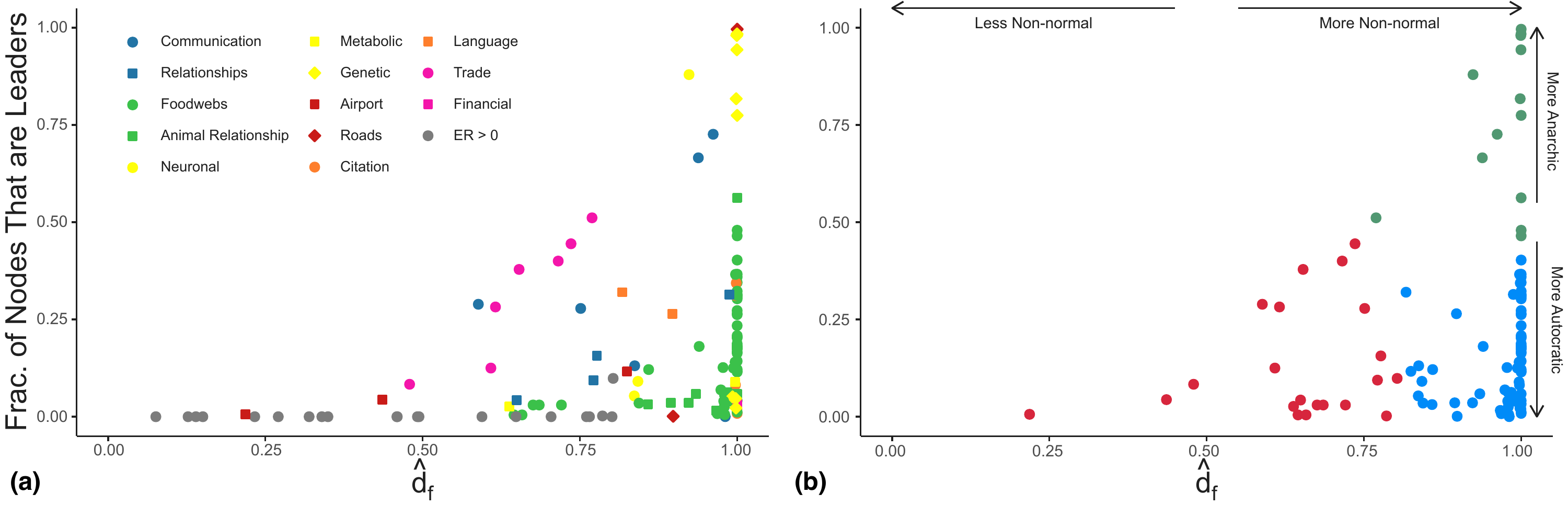}
		\caption{\textbf{The {hierarchical organization in} empirical networks{: oligarchy vs. anarchy.}} The fraction of {leader} nodes ({nodes without} outgoing links) in each network versus the level of non-normality present in the network captured via its normalized Henrici departure from normality $\hat{d}_f$. The color and shape of the points indicate the {different} domains {where} the networks {belong to, with the same notation used in Fig. \ref{fig:entropy_rate_emp}. \textbf{(b)} {Equivalent to (a) but where the two quantities are clustered via a k-means approach with $k = 3$. We see a clear pattern {of} clusters: first,  the cases where those networks with a larger fraction of leader nodes which describes the different form of leadership ranging from autocratic to anarchic{; the} second clear grouping {distinguishes} the level of directionality of the networks captured via the non-normality measure.}
		}}
		\label{fig:leaders_ER}
	\end{figure*}
	
	\subsubsection{Emergence of leaders}
	\noindent
	The {observed monotonic behavior} in the {relative} entropy rate $\hat{h}$ across networks from numerous domains as the level of non-normality {increases and eventually} surpasses a certain threshold, implies that the {dynamics of the }processes taking place on these networks are becoming less random {as a consequence of the increasing directedness of the underlying structure}. {This behavior is a result of an increasing accumulation of mass within a given set of nodes who receive from, but do not contribute to, other nodes in the network (terminal SCC). }{However, the abundance of cases {in which a complete collapse of the} entropy rate {occurs} indicate that the {accumulation} has {in fact reached a critical state} {whereby} the mass has {been} accumulated on single nodes, where it is trapped forever. 
	{In this paper,} we focus on how these leaders emerge {as a consequence of} the non-normality underlying the network structure. 
	{As such} {we pose the following question}:~are they created at random {while the {non-normality} increases}? To respond to these issues, we have measured the number of leaders {in each empirical network}, as shown in Fig.~\ref{fig:leaders_ER}. It can be immediately noticed that {(almost) all the scrutinised} empirical networks {which} have $\hat{h}\neq 0$ {have no} leaders in their structure. Furthermore, we observed a collapse of the entropy rate {across the ensemble of networks from all domains}. {Such} results {validates our belief} that {the} emergence of leaders is not an unorganized behavior, but {on} the contrary, {such occurrence is simultaneous} once a given threshold of non-normality is reached.
	
	{Figure~\ref{fig:leaders_ER}(a) gives us {further information regarding} the percentage of leaders in the real-world networks. Inspired by the term leader we use in this paper, we will {characterise} the networks accordingly to the relative number of leaders per network {(similarly to those found in Ref.~\cite{Johnson2020digraph})}. If the accumulation of mass {in the steady-state within a given terminal structure} is shared between several nodes, then we view the network as having an \emph{oligarchic} structure; this is the case, for instance, for all the networks with $\hat{h}\neq 0$. On the contrary, when the leaders emerge {with increasing non-normality}, the organization can be considered \emph{autocratic} or \emph{anarchic}, respectively, if the overall fraction of leader nodes is low or high. For example, most empirical networks belonging to the neuronal, animal relationship, social relationship, etc., subdomains tend to have a strong autocratic structure. On the other side, networks {such as} metabolic and genetic ones are highly anarchic with a star-like shape. Other domains {such as} food webs and communications networks, {however}, may vary between an autocratic to a more anarchic organization.} {This classification is further demonstrated via a {k-means} clustering approach~\cite{Jain2010} in Fig.~\ref{fig:leaders_ER}(b) which naturally finds the three classes of organizational structure described above.}
	
	\subsubsection{Hierarchical Structure}\label{sec:hierarchical}

	\begin{figure*}[t!]
		\centering
		\includegraphics[width= 0.9\textwidth]{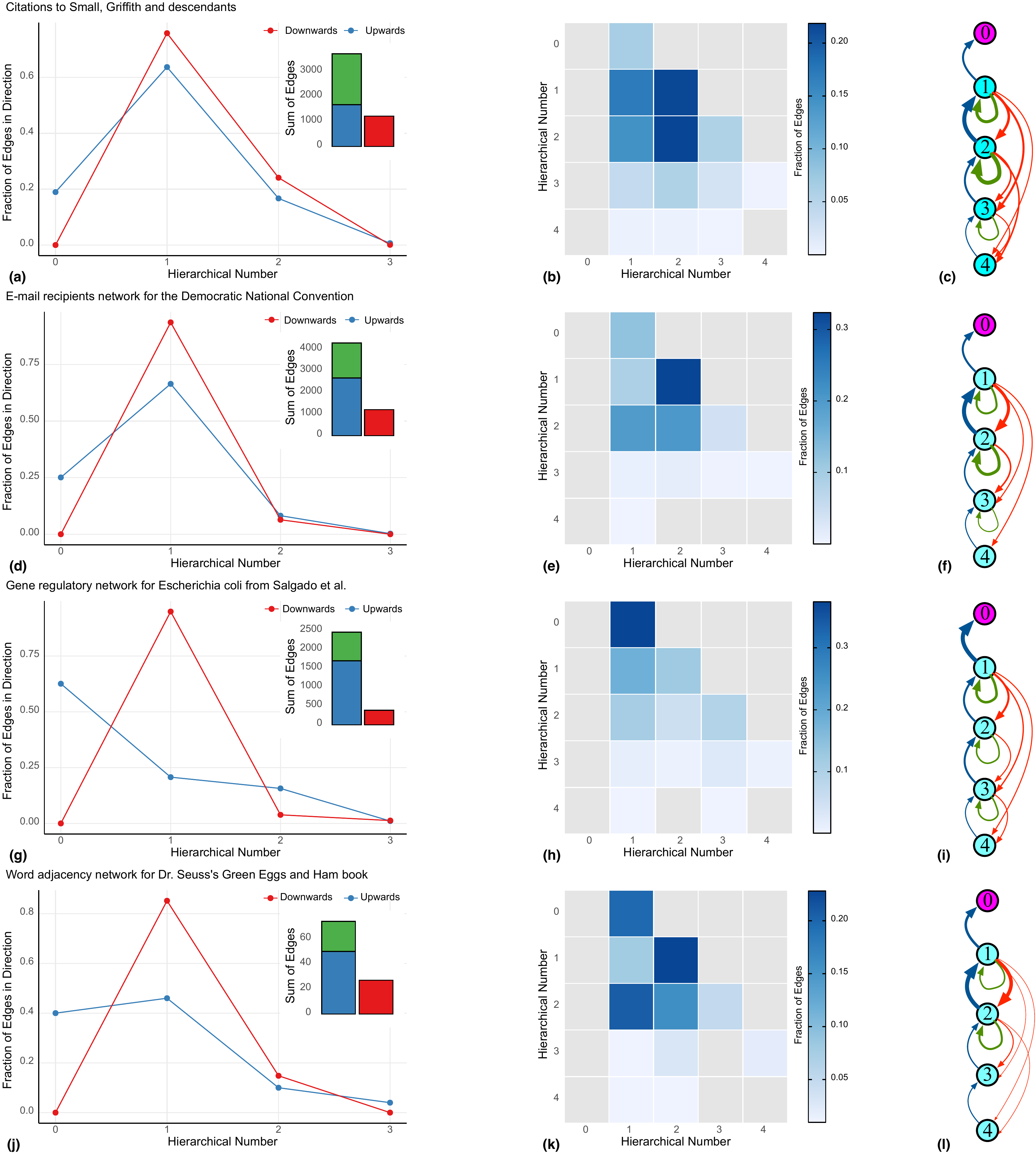}
		\caption{
			\textbf{Hierarchical structure of real-world networks.} \textbf{(a)} The fraction of edges to each hierarchical level {entering (respectively leaving)} from a lower level, blue lines, and {(respectively)} from a higher {hierarchical level}, red lines. {The i}nset plot shows the total weight of edges which are upwards (blue), downwards (red), and between hierarchies (green) in the case of the citations graph to Small \& Griffith (2001). \textbf{(b)} The fraction of edges between each hierarchical level is shown where we see that a large fraction of edges either move up hierarchical levels or else stay within their own hierarchy. \textbf{(c)} Illustrative schematic of the network's hierarchical structure. Equivalent plots for \textbf{(d)}-\textbf{(f)} the email network at the Democratic National Convention (2016), \textbf{(g)}-\textbf{(i)} the gene regulation network of the Escherichia coli, and \textbf{(j)}-\textbf{(l)} the word association network from the novel Green Eggs and Ham. {In all the cases a significantly higher number of upwards edges is observed associated to nodes of hierarchical level immediate to the leader nodes, the entourage nodes. It can also be noticed that the downwards red links, in fewer numbers tend to redistribute the flow from higher to lower hierarchical level. And last, more green links are associated to the entourage set of nodes, yielding a rich-club effect.}}
		\label{fig:emp_hierarchy}
	\end{figure*}
	\noindent
	{The results presented so far {consider} only a special subset of nodes, the leaders, without any {discussion} {as to} how they can be related to the other nodes of the network.} Motivated by the identification of {emergent behavior} {across} our ensembles of non-normal empirical networks we now proceed to consider how the remainder of the topology {describing} these networks is shaped in relation to their leaders. {Specifically, we} consider a ranking of the nodes} {in relation to their position in the} hierarchical {structure underlying the leader nodes}. {In this} sense {each} node {has} a level of importance based upon their proximity to a leader. We first identify each of the leaders before searching for shortest paths {originating} from these nodes~\cite{newman2010networks} {to each other node in the network}. {This} results in {each of} the network's {constituents} having a hierarchical label based upon their minimum distance to a leader as outlined in {the Supplementary Material (SM)}. {The resulting rankings} is {such} that leaders have a hierarchical label of zero, their direct neighbors a label of one, and so on.
	
	With the labels obtained for all nodes in {each} network, we proceed to determine the types of {relationship} facilitated by each edge{, schematically visualised in Fig. \ref{fig:emp_hierarchy}. Firstly,} there are \textit{ascending edges} (blue) that are aimed towards, and thus contribute to {directing the flow to the} leader. {In contrast,} we denote \textit{descending edges}~(red) {as those} that shift the flow from nodes of higher hierarchy, so nearer to the leader, towards those {of} lower hierarchy. Lastly, those edges which are between two nodes with the same hierarchical level, entitled \textit{{neutral} edges}~(green), thus keeping mass at a certain level. {We highlight} that by definition, neither descending nor {neutral} edges can originate from leader nodes.} 
	
		\begin{figure*}[t!]
		\centering
		\includegraphics[width=\textwidth]{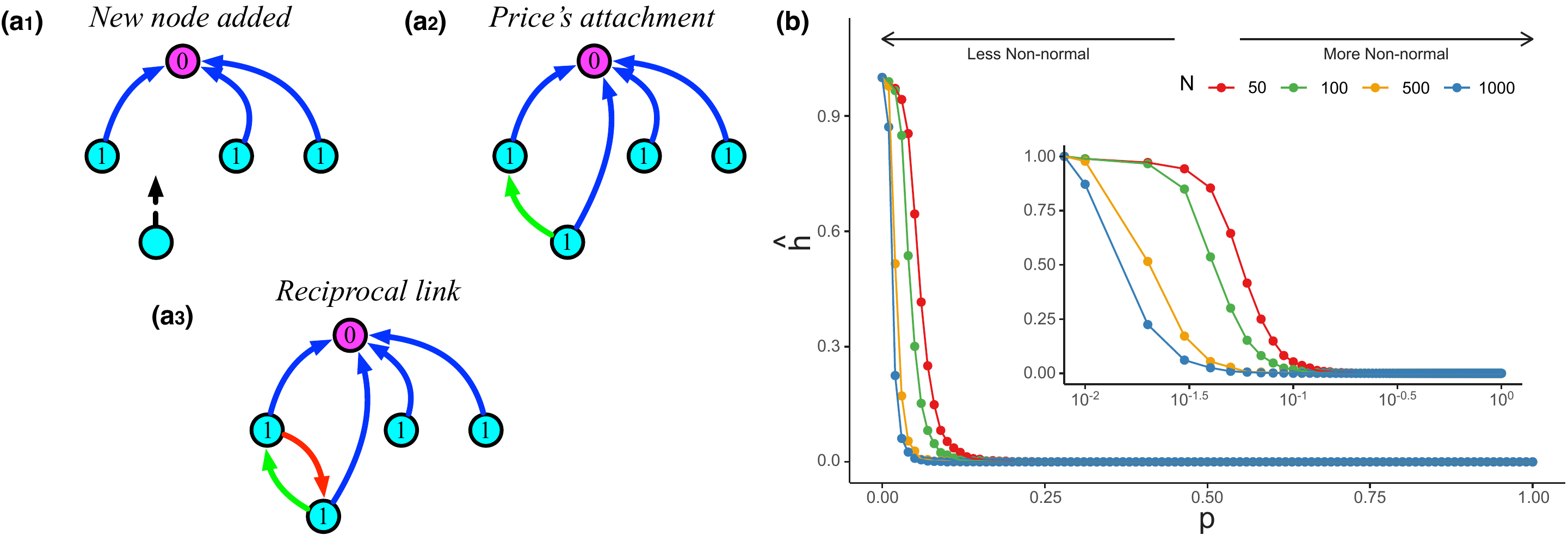}
		\caption{\textbf{Entropy rate of the generic random-walk on synthetic networks.}  \textbf{(a)} Schematic demonstrating {of the null model at multiple points in its growth process}. \textbf{(b)} Normalized entropy rate for the {null model} for a number of different network sizes {with $m = 3$} and threshold values $p$, {with the inset demonstrating} the same {quantities} with the horizontal axis on a log-scale, each point shown is the average over $10^4$ realizations.}
		\label{fig:entropy_rate_synthetic}
	\end{figure*}	
	
	{The organization of the nodes following the recipe described above} is now explicitly considered for a set of empirical networks from different domains - the citation network to the Small \& Griffith {paper up to the year} 2001~\cite{pajek_data}, the email network from the Democratic National Convention in 2016~\cite{konect}, the E. coli's gene regulatory network~\cite{Salgado2013}, and lastly the network of concatenated words in Dr.~Seuss's {novel} \textit{Green Eggs and Ham}~\cite{johnson2017looplessness}. For an extended analysis of the entire dataset of networks, the interested reader may refer to {SM}. In each of the leftmost {panels}, we consider
	the fraction of (weighted) blue and red edges {that correspondingly enter or leave the hierarchical levels indicated in the {horizontal} axis}. The inset histograms demonstrate the total sum of edges for each of the three edge types above. We see in each case {a common structural pattern} {(the same pattern can be observed for the {most of the empirical} networks in SM) where{, in particular,} a {considerably} larger fraction of blue and green edges {in comparison} to red ones indicates that the flow in the ascending hierarchy is considerably higher than {that in} the opposite direction.} The center {panels} provide an insight into the exact hierarchical structure by indicating the fraction of edges between each hierarchy where we again notice a {considerably} larger proportion of blue (represented here by the upper triangular elements) and green (those along the diagonal) edges {compared to red ones (in the lower triangular part)}. This {analysis provides} quantitative {evidence} that the structure {of empirical networks} is such to prove beneficial to {those nodes closer to the pinnacle of the hierarchical structure}. The right side schematics provide an illustrative indication of the type of structure present in the network. {Remarkably, it can be noticed here that nodes belonging to the hierarchical levels right after the leader have, on average, a high concentration of incoming ascending edges and neutral self-loops (this can also be {further} noticed in the SM). {This occurrence is suggestive of} a rich-club like effect~\cite{Colizza2006}. As we {shall} show {in the sequel}, both these features {prove to be} ultimately beneficial to the leaders and the nodes immediately associated with them, that we {denote} here as \textit{entourage} nodes.}
		
	\subsection{Network generation models}
	\noindent
  
	{After {the preceding} systematic empirical study of the hierarchical properties of the real-world networks,} we {now} consider mechanistic models {with the aim of shedding light} --- {both} analytically and through simulation --- {upon} {the possible mechanisms that relate {a network's} non-normality 
	and {the} {corresponding} emergence of leader nodes}. {With this motivation we proceed to} propose a novel model which {is based on the structural features observed} within the empirical networks {shown} in Sec.~\ref{sec:empirical} 
	{ and, {most importantly, that} can} capture {the emergence of leader nodes}. 
	
	The generation mechanism can be seen as organized in two stages: first, we create a network using the {renowned} Price's model originally used to model the emergence of a citation network~\cite{price1965networks, barabasi1999emergence}. According to this recipe, 
	{at} each time step, a new node~$j$ creates {$m$ directed edges} to~$m$ {(distinct)} already present nodes where the likelihood of joining to {a} node~$i$ is proportional to {its} in-degree. Importantly in the case of $m > 1$ this network immediately entails two interesting features - it is an exact DAG with one leader node (the first to appear) and also exhibits a hierarchical structure that may contain each of the three types of edges described in Sec.~\ref{sec:hierarchical}. In fact, although the wiring of new incoming {nodes} is more likely to {be towards} the nodes with high in-degree {(and generally closer to the leader)} in the case of $2$ or more links, connections to nodes with the same or lower hierarchical level {may} {also} occur. In order to control variations in the level of non-normality, we move to the second stage whereby we consider creating reciprocal edges of those generated in the first stage, similarly as done in Ref.~\cite{asllani2018structure}. The reasoning behind considering the reciprocal links is to capture the entire spectrum of behaviors from symmetric to DAG networks which would be otherwise impossible. An important observation here is that, if the distribution of reciprocal edges is uniform, this will lead to a larger number of edges from the seed node in the Price's model (who generally {has a large number of incoming edges}). This goes against the {hierarchical structure} observed in the empirical networks in Fig.~\ref{fig:emp_hierarchy}. To deal with such an issue, we will distribute the reciprocal edges according to a fitness model inspired by the well-known Bianconi-Barab\'asi model~\cite{Bianconi2001}. {So we} generate a reciprocal edge~$i$~$\to$~$j$ with probability {proportional to}~$1/k_i^{\text{out}}$, {such that an edge is included if this quantity surpasses a certain threshold $p$} with which the level of non-normality may be varied, thus decreasing the role of a node's importance in the first stage, and maintaining the distribution of hierarchical edges observed in the empirical networks.
	{Note in this case for $p = 0$ all reciprocal edges are drawn resulting in a symmetric network while $p = 1$ implies an exact DAG.} 

	\begin{figure*}[t!]
		\centering
		\includegraphics[width= \textwidth]{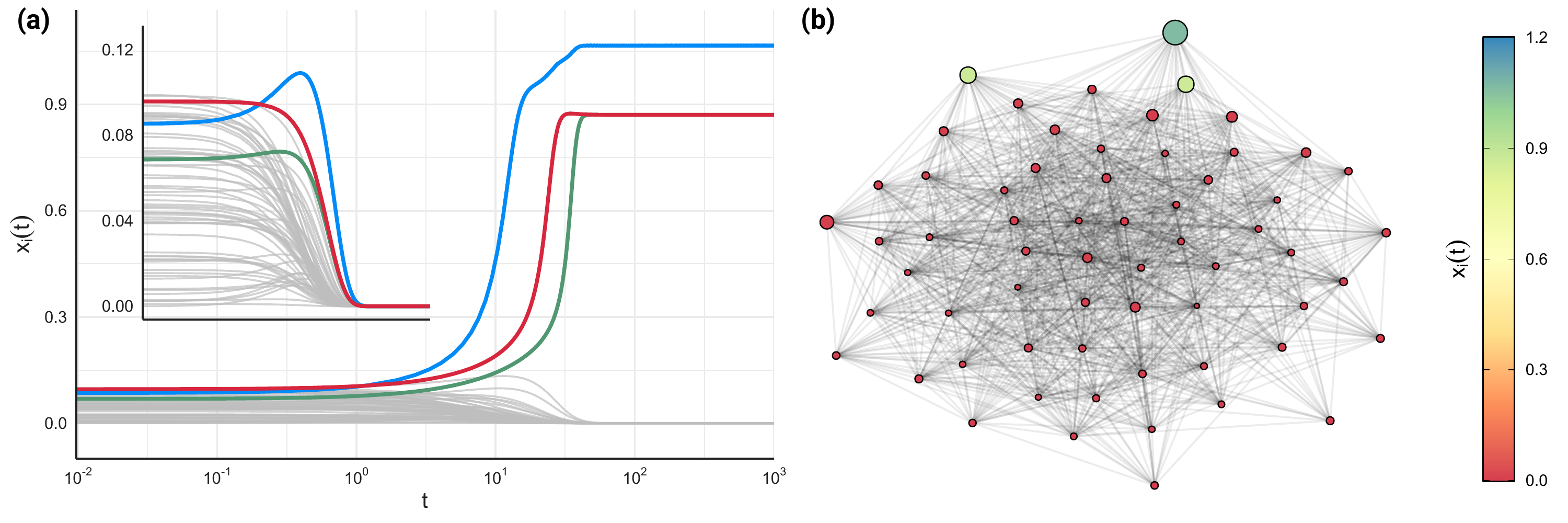}
		\caption{\textbf{Dynamical processes taking place on a dominance network of Japanese Macaque {monkeys}.} \textbf{(a)} The time evolution of $x_i(t)$ for both the Allee model {(solid)} and the linearised {version of the} model {(inset)}. In spite of the network only having one leader {(whose behavior is shown by the blue lines)} there are three nodes which survive in the Allee model shown by the colored lines. Note the same three nodes show the strongest transient growth in the {linearised} model. \textbf{(b)} The network {representation} with direction of edges omitted to improve clarity where the size of the nodes are inversely proportional to their outdegree {(the leader is thus the largest)} while their color represents the node's final {density}.}
		\label{fig:dynamical}
	\end{figure*}
	
	A schematic demonstration of this model  {at the multiple growing stages} is provided in Fig.~\ref{fig:entropy_rate_synthetic}(a) while simulations of such networks {with $m = 3$} and their corresponding normalized entropy rate as a function of {parameter} $p$ are shown in Fig.~\ref{fig:entropy_rate_synthetic}(b). {To validate our model, we compare {properties from ensembles of synthetic graphs} to the ground truth data where the entropy rate transition occurs. It can be observed in the inset of Fig.~\ref{fig:entropy_rate_emp} that the model fits very well{, in spite of its relative simplicity,} {capturing very well} the emergent behavior of the leader nodes.} {Notice that simpler generation models that, although not able to describe real data's behavior, give a good intuition in the relationship between non-normality and the ER are considered in Supplementary Material.} 
	
	\subsection{{The r}ole of leaders in {dynamical} processes}	
	\noindent
	{So far, we} have demonstrated {that empirical networks are characterized by }a rich structure {which {apparently evolves across ensembles of networks with increasing non-normality, culminating in the occurrence of leader nodes.} Although the importance of strongly directed hierarchies {has been shown to be a} signature of many complex systems in nature, and the decisive role of non-normality {in the dynamics} has {similarly} been {highlighted on} several occasions \cite{neubert_caswell_1997, Trefethen2020, Asllani2018PRE, asllani2018structure, baggio2020efficient}, {to} the best of our knowledge, {the ubiquitous occurrence of} an abundance of leader nodes {within} natural systems and in particular {their apparent} emergence in relation to a global measure such as non-normality {has yet to be discussed}. Consequently, it is important to illustrate, from a holistic point of view, the role of said leaders in the} the dynamical processes taking place on such systems. 
	{{With the aim of providing this} exemplification {in a generic manner}, we consider a competition process where $N$ identical individuals, among whom  a leader {exists}, compete for energy or mass (or resources in a more general term), {measured} by $\mathbf{x} = [x_1,x_2,\dots,x_N]$, which flows through the connections between the individuals} encoded by {the adjacency matrix} $\mathbf{A}$. {To keep the model simple, we consider here a bistable {dynamical system} with two possible (stable) states that each individual $i$ can have, {namely} it can {either} go extinct, {where} $x_i=0$ or survive {in the} case {$x_i > 0$}. {We describe this process mathematically via} the following system of diffusively coupled equations
	\begin{equation}
	\dot{x}_i = rx_i(1 - x_i)\left(\frac{x_i}{A} - 1\right) + D\sum_{j = 1}^{N} \mathcal{L}_{ij} x_j,\;\; \forall i
	\label{eq:allee_model}
	\end{equation}
	where $x_i$ describes the density of the $i{\text{-th}}$ species, $\mathcal{L}_{ij}~=~A_{ij}/k_j^{out}~-~\delta_{ij}$ are the entries of the random walk Laplacian matrix, $r$ is the reproductive rate, $D$ is the diffusion coefficient, {and $A$ is a parameter which allows the introduction of an unstable state, necessary for the bistability.} Notice that the transport operator used here is the mean-field equivalent of the random walk process considered {throughout} for the entropy rate~\cite{newman2010networks}. {This} model resembles {that} used to describe a phenomenon known in ecology as the Allee effect which describes the principle that undercrowding or a small density of a species' population decreases the likelihood of said species surviving~\cite{Allee1932}. {Recently it has been shown that,} for the case of symmetric networks, when the initial densities are {small, i.e., $0<x_i(0) \ll 1, \,\forall \, i$} each {individual} becomes extinct. {Conversely for} a non-normal system the behavior can result in some species surviving with some {equilibrium} density {$\mathbf{x_e\neq 0}$}~\cite{Asllani2018PRE}. 
	
	{{To provide some understanding of} the particular role of leaders }we {simulate this process upon an empirical} network constituted by a group of female Japanese macaque monkeys where the edges of the network represent dominance interactions between two animals~\cite{Takahata1991}{, and {in which} a single leader exists}. Figure \ref{fig:dynamical}(a) shows the {evolution of the mass of each node}~$x_i(t)${, through} the solid lines and we see the leader (blue) survives along with two other nodes (red and green) who are at the next level of hierarchy within the network. The network itself {(direction of edges omitted)} is shown in Fig.~\ref{fig:dynamical}(b) and we can observe the proximity of the nodes who survive both with one another and also the leader. {The outcome of these} dynamics provides an indication as to how the {specific} hierarchical {ranking} of {the individuals within the }network can be a benefit not only for the leaders but also those who position themselves {in close proximity} to said leaders.
	{In order to further comprehend} this phenomenon we {consider} the linearised model whose evolution is governed by the system of equations $\dot{x}_i= -rx_i + D\sum_{j = 1}^{N} \mathcal{L}_{ij} x_j$
	where {$r$} is the {decay}
	{(death)} rate and $D$ is the diffusion coefficient as before. The {evolution of the system in this case is presented{, with equivalent initial conditions to the non-linear case,} by} the inset of Fig.~\ref{fig:dynamical}(a). {Now }this {simplified} system has a unique fixed point $\mathbf{x_e} = \mathbf{0}$, {that ultimately defines the final outcome also. Nevertheless, since the survivability of each node depends on the balance of mass received and released per unit of time, for the case when the decay is slow compared to the diffusion rate, $r\ll D$, it might occur that nodes that have a high positive balance will initially accumulate {a larger quantity of} mass {in comparison to} the other {nodes} {before} {eventually} losing it {in} the asymptotic regime. This behavior is known in  theliterature as} transient growth~\cite{Trefethen2020} {and characterizes a large number of real systems~\cite{asllani2018structure}. However, such transient growth can turn into an instability mechanism when we deal with nonlinear systems, pushing the system unexpectedly far from the steady-state predicted from the linear analysis. This is the case for the individuals who survive in our scenario. In particular, {having the role of a} leader {implies} that the flow balance will always be large and positive, constituting a major benefit for the individual under consideration.}

	\section{Discussion and Conclusions}
	\noindent
	In this paper we have studied the architecture {of the hierarchical structures underpinning} {a large {collection} of }empirical networks through the lens of {the emerging} {leader nodes}. {Based on a tool borrowed from information theory} --- the {entropy rate} --- we conducted a study {aimed to} quantify {the amount of randomness {underlying} the distribution of the} equilibrium state {for} {a random walk }process {occurring} {on top of the network under analysis.} In particular, {we have related the configuration of equilibrium states to the polarization of the network structure by quantifying the {latter} through a global measure of the network, {namely} the graph} non-normality \cite{Trefethen2020,Asllani2018PRE,asllani2018structure}. Considering such a setting, we observed a remarkable property for the entropy rate{, which universally applies to all the real-world networks, {specifically, it} decreases monotonically while the {level of non-normality describing} the networks increase.} 
		
	One surprising result found to be particularly interesting is {how} the entropy rate {exhibits an abrupt collapse across the ensemble of networks once their} non-normality level succeeded a certain threshold. {We show that t}his phenomenon is immediately related to the  emergence of {leader }nodes{, namely {those} with only incoming (or only outgoing) edges, {across the networks}. In fact, these nodes are not present {for  {the} more normal networks} and {instead} appear {once a certain threshold has been reached.}} With these nodes identified, we proceeded to {obtain a} {hierarchical ranking of} the other nodes based upon their distance from the leaders. We used the resulting orientation to {identify three categories of links: {those that direct} the flow towards the leaders, those that redistribute the mass from the leaders to other nodes with lower hierarchy, and {lastly}, the intermediate edges that link nodes of the same hierarchical level. Ubiquitously, the links that ``feed'' the leaders are a considerable majority compared to both other categories. Based on these observations, we developed a null model for the generation of non-normal networks with the aforementioned topological properties, capturing the ground truth relationship between the entropy rate and the network {non-normality}, particularly the discontinuous transition {behavior} {which} yield the leader nodes.} {This {apparently ubiquitous} behavior {across domains} is characteristic of {those found in} first-order phase transitions~\cite{Sole1996}}.
	
	{The leader nodes, {either {sinks or sources} depending upon interpretation,} {can} eventually {prove} crucial in different scenarios. Possible examples can be {found} in ecology, e.g., the dominance hierarchies among individuals of animals (sink nodes that receives the ``benefits'' from other members) or food webs (sink nodes where the biomass accumulates); control engineering, e.g., the master-slave coupling of oscillators (source node, the ``master'' node who impose the oscillating frequency and phase); social interactions, e.g., contagion dynamics (source node {that} seeds the infection), etc. However, the leaders' role in collective behavior, particularly concerning the underlying non-normal dynamics, has been neglected so far. 
	{The present} paper briefly illustrates this role in {the case of} a simple competitive dynamics between individuals occurring on an empirical dominance hierarchical network. We show that the privileged status that a leader node has is related to the fact that it absorbs the flow 
    without the constraint of releasing. Of particular importance is the balance of incoming and outgoing flux that a node has, which {results} in {an} advantage even for the entourage nodes, {those} immediately {connected} to the {leaders}.}
	
	{{Based on the apparent} ubiquity of leaders in real-world systems, {we are confident that our finding will trigger future exciting research directions, {and contribute to}  better understanding} how different dynamical systems are affected by this emergent phenomenon. 
	
	\section{Methods}
	\subsection{Random walk process and its entropy}
	\noindent
	The random-walk process considered in this article describes some quantity which we refer to as mass that {is transported} between nodes such that at each time its particles move from node $j$ to one of its neighbors $i$ with {transition} probability 
	\begin{equation}
	T_{ij} = \begin{cases}
	\dfrac{w_{ij}}{k_j^{\text{out}}} & k_j^{\text{out}} \ne 0, i \ne j\\
	1 & k_j^{\text{out}} = 0, i = j \\
	0 &\text{otherwise,}
	\end{cases}
	\label{eq:trans_matrix}
	\end{equation} 
	where $k_j^{\text{out}} = \sum_i w_{ij}$ is the outdegree of node $j$. {Note that following this definition the mass cannot leave the nodes without outgoing edges.} We consider, as in \cite{gomez2008entropy, burda2009localization}, the probability $q_j(t)$ that the random walker who represents a unit of mass is present at node $j$ at time $t$ such that the vector $\mathbf{q}(t) = \left[q_1(t), q_2(t), \cdots, q_N(t)\right]$ describes the proportion of mass with each node at time $t$ (with $\sum_j w_j(t) = 1$). The dynamics of this system is thus given by $\mathbf{q}(t+1) = \mathbf{T} \, \mathbf{q}(t)$. In general we are concerned with the long-time behavior of these systems, i.e., the stationary distribution $q_j^* = \lim_{t\to\infty} q_j(t)$, the existence of a unique occurrence of this distribution is very much dependent on the structure of {$\mathbf{T}$} which is itself determined by $\mathbf{A}$. {Finally the} entropy rate, {which represents the amount of information required to describe the diffusion process in question~\cite{cover2005information,gomez2008entropy},} {is given by} $h = - \sum_{i,j} T_{ij} \times q_j^* \ln(T_{ij})$. Importantly one may {notice} that in the case of (sink) {leader} nodes, this process results in an entropy rate of value zero as in this case all the mass is accumulated in these nodes and} the only  {contributing term is due to the leader node $j$ with $T_{jj}=1$.}

	\begin{acknowledgements}\noindent
	This work was supported by Science Foundation Ireland Grant No. 16/IA/4470, No.~16/RC/3918, No.~12/RC/2289P2 and No.~18/CRT/6049 (J.D.O'B., K.A.O., J.P.G., and M.A.). We acknowledge the DJEI/DES/SFI/HEA Irish Centre for High-End Computing for the provision of computational facilities. 
	\end{acknowledgements}

	\bibliographystyle{Science}
	\bibliography{NN_bib}

\begin{thebibliography}{100}

\bibitem{newman2010networks}
M.~Newman, {\it {Networks: An Introduction}\/} (Oxford university press, 2010).

\bibitem{boccaletti2006complex}
S.~Boccaletti, V.~Latora, Y.~Moreno, M.~Chavez, D.~U. Hwang, {\it Phys. Rep.\/}
  {\bf 424}, 175 (2006).

\bibitem{johnson2017looplessness}
S.~Johnson, N.~S. Jones, {\it Proc. Natl. Acad. Sci. USA\/} {\bf 114}, 5618
  (2017).

\bibitem{Johnson2020digraph}
S.~Johnson, {\it J. Phys. Complexity\/} {\bf 1}, 015003 (2020).

\bibitem{kawakatsu2020emergence}
M.~Kawakatsu, P.~S. Chodrow, N.~Eikmeier, D.~B. Larremore, {\it arXiv
  arXiv:2007.04448\/}  (2020).

\bibitem{asllani2018structure}
M.~Asllani, R.~Lambiotte, T.~Carletti, {\it Sci. Adv.\/} {\bf 4}, 1 (2018).

\bibitem{Asllani2018PRE}
M.~Asllani, T.~Carletti, {\it Phys. Rev. E\/} {\bf 97}, 1 (2018).

\bibitem{Trefethen2020}
L.~N. Trefethen, M.~Embree, {\it {Spectra and Pseudospectra}\/} (Princeton
  University Press, 2020).

\bibitem{neubert_caswell_1997}
M.~G. Neubert, H.~Caswell, {\it Ecology\/} {\bf 78}, 653–665 (1997).

\bibitem{Ravoori2011}
B.~Ravoori, {\it et~al.\/}, {\it Phys. Rev. Lett.\/} {\bf 107}, 1 (2011).

\bibitem{Hennequin2012}
G.~Hennequin, T.~P. Vogels, W.~Gerstner, {\it Phys. Rev. E\/} {\bf 86}, 1
  (2012).

\bibitem{pilgrim2020organisational}
C.~Pilgrim, W.~Guo, S.~Johnson, {\it Sci. Rep.\/}  (2020).

\bibitem{Muolo2019}
R.~Muolo, M.~Asllani, D.~Fanelli, P.~K. Maini, T.~Carletti, {\it J. Theor.
  Biol.\/} {\bf 480}, 81 (2019).

\bibitem{NN_stoch}
S.~Nicoletti, {\it et~al.\/}, {\it Chaos\/} {\bf 29}, 083123 (2019).

\bibitem{baggio2020efficient}
G.~Baggio, V.~Rutten, G.~Hennequin, S.~Zampieri, {\it Sci. Adv.\/} {\bf 6},
  eaba2282 (2020).

\bibitem{cover2005information}
T.~M. Cover, J.~A. Thomas, {\it {Elements of Information Theory}\/} (John Wiley
  \& Sons, 2005).

\bibitem{gomez2008entropy}
J.~G{\'{o}}mez-Garde{\~{n}}es, V.~Latora, {\it Phys. Rev. E\/} {\bf 78}, 1
  (2008).

\bibitem{ER_crowd}
T.~Carletti, M.~Asllani, D.~Fanelli, V.~Latora, {\it Phys. Rev. Research\/}
  {\bf 2}, 033012 (2020).

\bibitem{price1965networks}
D.~J. {De Solla Price}, {\it Science\/}  (1965).

\bibitem{Allee1932}
W.~C. Allee, E.~S. Bowen, {\it J. Exper. Zoology\/}  (1932).

\bibitem{masuda2017random}
N.~Masuda, M.~A. Porter, R.~Lambiotte, {\it Phys. Rep.\/} {\bf 716-717}, 1
  (2017).

\bibitem{Note1}
{An exemption occurs when each node of the terminal SCC has a single outgoing
  edge, namely forms a closed directed path. Network motifs such as closed
  directed paths are very rare in real-world networks~\cite {Milo_2002} mainly
  due to the inherent directedness of real networks~\cite {Dominguez_2014}}.

\bibitem{Note2}
Although they have a relatively small non-normality due to generally having
  reciprocal (albeit weighted) links, there are interestingly some airports
  from which passengers only arrive (sink nodes), {resulting in an} anomalous
  {decrease} in the entropy rate. We believe that this is because such systems
  are open (the passengers leave from other means {in the case of} sink
  airports), making them unsuitable for consistent analysis.

\bibitem{Jain2010}
A.~K. Jain, {\it Patt. Rec. Lett.\/} {\bf 31}, 651 (2010).

\bibitem{pajek_data}
{Pajek datasets}, \url{http://vlado.fmf.uni-lj.si/pub/networks/}.

\bibitem{konect}
{J. Kunegis}, {The KONECT project}, \url{http://konect.cc/}.

\bibitem{Salgado2013}
H.~Salgado, {\it et~al.\/}, {\it Nucleic Acids Research\/} {\bf 41}, D203
  (2013).

\bibitem{Colizza2006}
V.~Colizza, A.~Flammini, M.~A. Serrano, A.~Vespignani, {\it Nature Phys.\/}
  (2006).

\bibitem{barabasi1999emergence}
A.~L. Barab{\'{a}}si, R.~Albert, {\it Science\/}  (1999).

\bibitem{Bianconi2001}
G.~Bianconi, A.~L. Barab{\'{a}}si, {\it Europhys. Lett.\/} {\bf 54}, 436
  (2001).

\bibitem{Takahata1991}
Y.~Takahata, {\it The Monkeys of Arashiyama\/} pp. 124--139 (1991).

\bibitem{Sole1996}
R.~V. Sol{\'{e}}, S.~C. Manrubia, B.~Luque, J.~Delgado, J.~Bascompte, {\it
  Complexity\/}  (1996).

\bibitem{burda2009localization}
Z.~Burda, J.~Duda, J.~M. Luck, B.~Waclaw, {\it Phys. Rev. Lett.\/} {\bf 102}, 1
  (2009).

\bibitem{Milo_2002}
R.~Milo, {\it et~al.\/}, {\it Science\/} {\bf 298}, 824 (2002).

\bibitem{Dominguez_2014}
V.~Dom{\'{\i}}nguez-Garc{\'{\i}}a, S.~Pigolotti, M.~A. Mu{\~{n}}oz, {\it Sci.
  Rep.\/} {\bf 4} (2014).

\bibitem{ref55}
D.~Thieffry, A.~M. Huerta, E.~P{\'e}rez-Rueda, J.~Collado-Vides, {\it
  Bioessays\/} {\bf 20}, 433 (1998).

\bibitem{ref53}
M.~B. Gerstein, {\it et~al.\/}, {\it Nature\/} {\bf 489}, 91 (2012).

\bibitem{ref59}
C.~T. Harbison, {\it et~al.\/}, {\it Nature\/} {\bf 431}, 99 (2004).

\bibitem{ref62}
M.~C. Costanzo, {\it et~al.\/}, {\it Nucleic Acids Research\/} {\bf 29}, 75
  (2001).

\bibitem{ref57}
J.~Sanz, {\it et~al.\/}, {\it PloS one\/} {\bf 6}, e22178 (2011).

\bibitem{ref56}
H.~Salgado, {\it et~al.\/}, {\it Nucleic acids research\/} {\bf 41}, D203
  (2013).

\bibitem{ref39}
R.~M. Ewing, {\it et~al.\/}, {\it Molecular systems biology\/} {\bf 3}, 89
  (2007).

\bibitem{ref58}
E.~Gal{\'a}n-V{\'a}squez, B.~Luna, A.~Mart{\'\i}nez-Antonio, {\it Microbial
  informatics and experimentation\/} {\bf 1}, 3 (2011).

\bibitem{ref70}
N.~T. Markov, {\it et~al.\/}, {\it Proceedings of the National Academy of
  Sciences\/} {\bf 110}, 5187 (2013).

\bibitem{ref54}
D.~J. Watts, S.~H. Strogatz, {\it nature\/} {\bf 393}, 440 (1998).

\bibitem{ref7}
M.~Kaiser, C.~C. Hilgetag, {\it PLoS Comput Biol\/} {\bf 2}, e95 (2006).

\bibitem{ref66}
M.~Bota, L.~W. Swanson, {\it Journal of Comparative Neurology\/} {\bf 500}, 807
  (2007).

\bibitem{ref67}
C.~Carere, G.~F. Ball, J.~Balthazart, {\it Journal of Comparative Neurology\/}
  {\bf 500}, 894 (2007).

\bibitem{ref68}
R.~M. Cowell, K.~R. Blake, J.~W. Russell, {\it Journal of Comparative
  Neurology\/} {\bf 502}, 1 (2007).

\bibitem{ref69}
L.~Harriger, M.~P. Van Den~Heuvel, O.~Sporns, {\it PloS one\/} {\bf 7}, e46497
  (2012).

\bibitem{ref52}
H.~Jeong, B.~Tombor, R.~Albert, Z.~N. Oltvai, A.-L. Barab{\'a}si, {\it
  Nature\/} {\bf 407}, 651 (2000).

\bibitem{ref28}
S.~Johnson, Network data repository from various sources.,
  \url{https://www.samuel-johnson.org/data}. Accessed May, 26th, 2020.

\bibitem{ref43}
M.~A. de~Reus, M.~P. van~den Heuvel, {\it Journal of Neuroscience\/} {\bf 33},
  12929 (2013).

\bibitem{ref4}
R.~M. Thompson, C.~Townsend, {\it Ecology\/} {\bf 84}, 145 (2003).

\bibitem{ref5}
R.~M. Thompson, A.~R. Mcintosh, {\it Ecology Letters\/} {\bf 1}, 200 (1998).

\bibitem{ref9}
J.~Klaise, S.~Johnson, {\it Scientific reports\/} {\bf 7}, 1 (2017).

\bibitem{ref2}
J.~A. Dunne, {\it et~al.\/}, {\it PLoS Biol\/} {\bf 11}, e1001579 (2013).

\bibitem{ref13}
R.~M. Thompson, C.~Townsend, {\it Oikos\/} {\bf 108}, 137 (2005).

\bibitem{ref22}
J.~Memmott, N.~D. Martinez, J.~Cohen, {\it Journal of Animal Ecology\/} {\bf
  69}, 1 (2000).

\bibitem{ref23}
J.~Bascompte, C.~J. Meli{\'a}n, E.~Sala, {\it Proceedings of the National
  Academy of Sciences\/} {\bf 102}, 5443 (2005).

\bibitem{ref24}
R.~E. Ulanowicz, D.~Baird, {\it Journal of Marine Systems\/} {\bf 19}, 159
  (1999).

\bibitem{ref44}
B.~J. Cole, {\it Science\/} {\bf 212}, 83 (1981).

\bibitem{ref47}
T.~Grant, {\it Animal Behaviour\/} {\bf 21}, 449 (1973).

\bibitem{ref75}
R.~R. Christian, J.~J. Luczkovich, {\it Ecological modelling\/} {\bf 117}, 99
  (1999).

\bibitem{ref76}
L.~Goldwasser, J.~Roughgarden, {\it Ecology\/} {\bf 74}, 1216 (1993).

\bibitem{ref19}
M.~Huxham, S.~Beaney, D.~Raffaelli, {\it Oikos\/} pp. 284--300 (1996).

\bibitem{ref12}
A.~Ekl{\"o}f, {\it et~al.\/}, {\it Ecology letters\/} {\bf 16}, 577 (2013).

\bibitem{ref6}
J.~A. Dunne, R.~J. Williams, N.~D. Martinez, R.~A. Wood, D.~H. Erwin, {\it PLoS
  Biol\/} {\bf 6}, e102 (2008).

\bibitem{ref21}
K.~Havens, {\it Science\/} {\bf 257}, 1107 (1992).

\bibitem{ref51}
J.~A. van Hooff, J.~A. Wensing  (1987).

\bibitem{ref36}
N.~D. Martinez, {\it Ecological monographs\/} {\bf 61}, 367 (1991).

\bibitem{ref73}
J.~Link, {\it Marine ecology progress series\/} {\bf 230}, 1 (2002).

\bibitem{ref41}
F.~F. Strayer, M.~S. Cummins, {\it Dominance relations: an ethological view of
  human conflict and social interaction. Edinburgh: Livingstone\/}  (1980).

\bibitem{ref74}
P.~H. Warren, {\it Oikos\/} pp. 299--311 (1989).

\bibitem{ref20}
P.~Yodzis, {\it Journal of Animal Ecology\/} {\bf 67}, 635 (1998).

\bibitem{ref32}
R.~E. Ulanowicz, C.~Bondavalli, M.~Egnotovich, {\it Annual Report to the United
  States Geological Service Biological Resources Division Ref. No.[UMCES]
  CBL\/} pp. 98--123 (1998).

\bibitem{ref48}
Y.~Takahata, {\it The monkeys of Arashiyama. State University of New York
  Press, Albany\/} pp. 123--139 (1991).

\bibitem{ref49}
T.~Clutton-Brock, P.~Greenwood, R.~Powell, {\it Zeitschrift f{\"u}r
  Tierpsychologie\/} {\bf 41}, 202 (1976).

\bibitem{ref46}
M.~W. Schein, M.~H. Fohrman, {\it The British Journal of Animal Behaviour\/}
  {\bf 3}, 45 (1955).

\bibitem{ref50}
C.~C. Hass, {\it Journal of Zoology\/} {\bf 225}, 509 (1991).

\bibitem{ref77}
A.~Guhl, {\it Transactions of the Kansas Academy of Science (1903-)\/} {\bf
  71}, 379 (1968).

\bibitem{ref45}
D.~F. Lott, {\it Zeitschrift f{\"u}r Tierpsychologie\/} {\bf 49}, 418 (1979).

\bibitem{ref33}
M.~J. Bashaw, M.~A. Bloomsmith, T.~L. Maple, F.~B. Bercovitch, {\it Journal of
  Comparative Psychology\/} {\bf 121}, 46 (2007).

\bibitem{ref42}
V.~Krebs, Madoff feeder funds., \url{
  http://www.thenetworkthinkers.com/2009/02/madoff-feeder-funds.html}. Accessed
  September, 2nd, 2020.

\bibitem{ref31}
M.~Flandreau, C.~Jobst, {\it The Journal of Economic History\/} {\bf 65}, 977
  (2005).

\bibitem{ref30}
M.~De~Domenico, V.~Nicosia, A.~Arenas, V.~Latora, {\it Nature communications\/}
  {\bf 6}, 1 (2015).

\bibitem{ref60}
D.~A. Smith, D.~R. White, {\it Social Forces\/} {\bf 70}, 857 (1992).

\bibitem{ref61}
W.~De~Nooy, A.~Mrvar, V.~Batagelj, {\it Exploratory social network analysis
  with Pajek: Revised and expanded edition for updated software\/}, vol.~46
  (Cambridge University Press, 2018).

\bibitem{ref17}
E.~Garfield, Index of citation networks produced by analyses from the software
  {HistCite}, \url{http://www.garfield.library.upenn.edu/histcomp/index.html}.
  Accessed February, 19th, 2020.

\bibitem{ref16}
N.~P. Hummon, P.~Dereian, {\it Social networks\/} {\bf 11}, 39 (1989).

\bibitem{ref14}
A.~Schubert, {\it Scientometrics\/} {\bf 53}, 3 (2002).

\bibitem{ref26}
M.~Ley, {\it International symposium on string processing and information
  retrieval\/} (Springer, 2002), pp. 1--10.

\bibitem{ref34}
R.~Milo, {\it et~al.\/}, {\it Science\/} {\bf 303}, 1538 (2004).

\bibitem{ref35}
W.~De~Nooy, {\it Poetics\/} {\bf 26}, 385 (1999).

\bibitem{ref27}
B.~R. da~Cunha, {\it et~al.\/}, {\it Scientific reports\/} {\bf 10}, 1 (2020).

\bibitem{ref72}
B.~C. Boatwright, D.~L. Linvill, P.~L. Warren, {\it Resource Centre on Media
  Freedom in Europe\/}  (2018).

\bibitem{ref18}
J.~Leskovec, D.~Huttenlocher, J.~Kleinberg, {\it Proceedings of the SIGCHI
  conference on human factors in computing systems\/} (2010), pp. 1361--1370.

\bibitem{ref1}
L.~A. Adamic, N.~Glance, {\it Proceedings of the 3rd international workshop on
  Link discovery\/} (2005), pp. 36--43.

\bibitem{ref25}
Kaggle, Chess ratings - elo versus the rest of the world (2010).,
  \url{https://www.kaggle.com/c/chess/data}. Accessed August, 27th, 2020.

\bibitem{ref8}
J.~Coleman, E.~Katz, H.~Menzel, {\it Sociometry\/} {\bf 20}, 253 (1957).

\bibitem{ref10}
J.~Kunegis, ``{DNC emails co-recipients.'' KONECT, the Koblenz Network
  Collection (2016)}, \url{https://networks.skewed.de/net/dnc}. Accessed
  February, 19th, 2020.

\bibitem{ref64}
T.~Opsahl, P.~Panzarasa, {\it Social networks\/} {\bf 31}, 155 (2009).

\bibitem{ref3}
T.~N. for Research Core~Team, Transportation networks for research,
  \url{https://github.com/bstabler/TransportationNetworks}. Accessed August,
  8th, 2020.

\bibitem{ref29}
U.~S. F.~A. Administration, Air traffic control system command center extracted
  prior to 2010., \url{https://www.fly.faa.gov/}. Accessed September, 2nd,
  2020.

\bibitem{ref15}
L.~J. LeBlanc, E.~K. Morlok, W.~P. Pierskalla, {\it Transportation research\/}
  {\bf 9}, 309 (1975).

\bibitem{ref65}
T.~Opsahl, F.~Agneessens, J.~Skvoretz, {\it Social networks\/} {\bf 32}, 245
  (2010).

\bibitem{ref37}
F.~Munoz-Mendez, K.~Han, K.~Klemmer, S.~Jarvis, {\it Proceedings of the 2018
  ACM International Joint Conference and 2018 International Symposium on
  Pervasive and Ubiquitous Computing and Wearable Computers\/} (2018), pp.
  1015--1023.

\bibitem{ref71}
C.~Demetrescu, 9th dimacs implementation challenge - shortest paths,
  \url{http://archive.dimacs.rutgers.edu/Workshops/Challenge9/}. Accessed
  November, 22th, 2019.

\bibitem{ref38}
M.~J. Williams, M.~Musolesi, {\it Royal Society open science\/} {\bf 3}, 160196
  (2016).

\bibitem{ref63}
OpenFlights, Airport, airline and route data.,
  \url{https://openflights.org/data.html}. Accessed September, 2nd, 2020.

\bibitem{ref11}
J.~Zhang, S.~Pourazarm, C.~G. Cassandras, I.~C. Paschalidis, {\it 2016 IEEE
  55th Conference on Decision and Control (CDC)\/} (IEEE, 2016), pp. 789--794.

\end{thebibliography}
	
	\onecolumngrid
	\renewcommand{\figurename}{}
	\renewcommand{\tablename}{}
	\renewcommand{\thefigure}{Supplementary Figure \arabic{figure}}
	\renewcommand{\thetable}{Supplementary Table \arabic{table}}
	\setcounter{figure}{0}
	\setcounter{section}{0}
	\setcounter{subsection}{0}
	\renewcommand{\theequation}{S\arabic{equation}}
	\addtocontents{toc}{\protect\setcounter{tocdepth}{0}} 
	\renewcommand{\thesection}{Supplementary Note \Roman{section}}
	\newcolumntype{P}[1]{>{\centering\arraybackslash}p{#1}}
	\renewcommand{\dbltopfraction}{0.9}
	\renewcommand{\floatpagefraction}{.8}%

	\begin{center}
		\textbf{\large Supplemental Materials}
	\end{center}

	\section{Non-normality Metrics}
	
	Throughout this work we focus on directed, weighted graphs described by the $N \times N$ adjacency matrix $\mathbf{A}$, which has elements $w_{ij}$ describing the weight of an edge from node $j$ to node $i$. In order to quantify the level of non-normality present in a given network we make use of two measures from matrix theory. The first of these is known as the Henrici departure from normality $d_F(\mathbf{A}) = \sqrt{||\mathbf{A}||^2_F - \sum_{i = 1}^{N} |\lambda_i|^2}$, where $||\cdot||_F$ describes the Frobenius norm and $\lambda_i$ represents the eigenvalues of the matrix~\cite{Trefethen2020}. As this quantity does not have a natural scale we instead consider the normalized Henrici departure from normality~$\hat{d}_F(\mathbf{A}) = d_F(\mathbf{A})/||\mathbf{A}||_F$, which varies from the extreme values for a symmetric network ($\hat{d}_F(\mathbf{A}) = 0$) to the case of an exact DAG ($\hat{d}_F(\mathbf{A}) = 1$). The second measure of non-normality considered in the article is known as the unbalance $\Delta$ between the number of entries in the upper and lower triangular elements of the adjacency matrix such that $\Delta = |K^< - K^>|/(K^< + K^>)$ where $K^< = \sum_{i < j}\mathbf{\tilde{A}}$, $K^> = \sum_{j < i}\mathbf{\tilde{A}}$, and $\mathbf{\tilde{A}}$ represents a relabeled version of the original adjacency matrix  obtained via an optimization procedure. A heuristic strategy looks for the matrix which maximizes the unbalance between its upper and lower triangles and the search space of this procedure is navigated through simultaneously swapping two randomly picked rows and their two corresponding columns of the original adjacency matrix. The heuristic implemented in this case is a simulated annealing, similar to  \cite{asllani2018structure}. Its output should approximate the closest the network structure may be to a DAG in one of the triangles of the resulting matrix. The two metrics are then shown in~\ref{fig:henrici_unbalance} in the case of the 124 empirical networks used in this study (the exact numeric quantities can be found in the~\ref{sec:data}). It may immediately be seen that a strong positive correlation exists between the two quantities demonstrating their usefulness in describing the level of non-normality present in a network and the pervasive nature of this feature among empirical systems. This relationship has previously been commented on for a smaller collection of networks in Ref.~\cite{asllani2018structure} thus the analysis provided here gives further evidence of the finding.

	\section{Synthetic Network Models}
	
	\subsection{An Exactly Solvable Model: The Chain Network}
	
	We now consider a linear graph, {in particular, a unweighted unidirectional} \textit{chain network} that has {been complemented with backward loops of weight $\epsilon$. The latter will be the control parameter with which we can tune the level of non-normality of our toy network. A similar network model has been also considered in~\cite{NN_stoch,baggio2020efficient}.} In this scenario, {we have a network that can vary from the case of a} {simple unidirectional} chain network {when} $\epsilon = 0$ and a fully symmetric version of the network {when} $\epsilon = 1$. An illustration of such a network is provided in~\ref{fig:chain_model}(a).
	
	We now consider the entropy rate of the generic random walk tkaing place upon this network given by
	\begin{equation}
		h = - \sum_{i,j} T_{ij} \times q_j^\star \ln(T_{ij}),
		\label{eq:ER}
	\end{equation}
	where each term is as in the main text. If we begin by noticing that this network's adjacency matrix is given by
	\begin{equation}
		A_{ij}=
		\begin{cases}
			1 & j = i + 1, i \ne N\\
			\epsilon & j = i - 1, i \ne 1 \\
			0 &\text{otherwise}
		\end{cases}	
	\end{equation}
	and thus the transition matrix of the random walk taking place on this system is given by
	\begin{equation}
		T_{ij}=
		\begin{cases}
			1 & j = 1, i = 2\\
			1 & j = N, i = N - 1 \\
			\dfrac{1}{1+\epsilon} & j = i + 1, i \ne N \\ 
			\dfrac{\epsilon}{1+\epsilon} & j = i - 1, i \ne 1 \\
			0 &\text{otherwise.}
		\end{cases}
		\label{eq:trans_chain}
	\end{equation}
	The stationary distribution of the random walk process occurring on the network is also required to determine the entropy rate and this is readily shown to be given by
	\begin{equation}
		q_j^\star = 
		\begin{cases}
			\dfrac{1}{1 + \epsilon^{N-2} + (1+\epsilon)\sum_{k = 0}^{N - 3}\epsilon^k} & j = 1\\
			\epsilon^{j - 2}(1+\epsilon) \, q_1^\star & 2 \le j \le N - 1 \\
			\epsilon^{N-2} \, q_1^\star & j = N
		\end{cases}
		\label{eq:stat_dist_chain}
	\end{equation}
	lastly substituting both Eqs.~\eqref{eq:trans_chain} and \eqref{eq:stat_dist_chain} into \eqref{eq:ER} allows one to calculate the entropy rate exactly as
	\begin{align}
		h &= -\sum_{j}q_j^\star \sum_{i} T_{ij} \log{(T_{ij})}, \nonumber \\
		&= -\sum_{j = 2}^{N-1} q_j^\star \left[\frac{\epsilon}{1 + \epsilon} \log\left(\frac{\epsilon}{1 + \epsilon}\right) + \frac{1}{1+\epsilon}\log\left(\frac{1}{1 + \epsilon}\right)\right], \nonumber \\
		&= -q_1^\star \left[\epsilon\log\left(\frac{\epsilon}{1 + \epsilon}\right) + \log\left(\frac{1}{1 + \epsilon}\right)\right] \frac{1 - \epsilon^{N - 2}}{1 - \epsilon}.
		\label{eq:chain_entropy}
	\end{align}
	
	\begin{figure}
		\centering
		\includegraphics[width=0.8\textwidth]{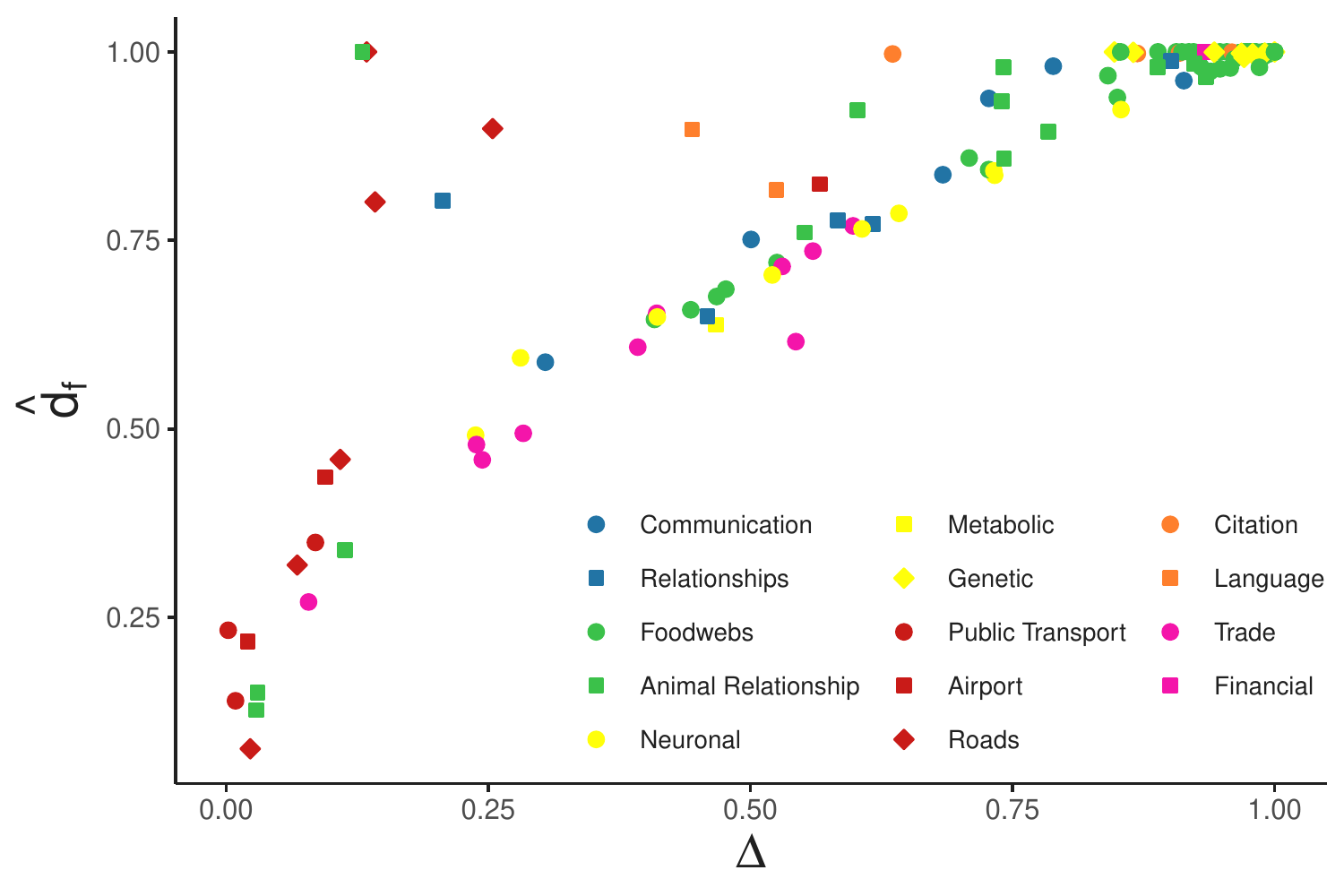}
		\caption{\textbf{The strong asymmetric and non-normal structure of empirical networks.} The normalized Henrici's departure from normality $\hat{d}_f$ versus the structural measure of asymmetry $\Delta$ for $124$ networks from a {large} range of domains is shown. We note the positive correlation between the two measures. {The data are grouped in $6$ domains represented by} the {same} color of the symbols which in turn are divided in several sudomains identifiable by different shapes.}
		\label{fig:henrici_unbalance}
	\end{figure}
	
	\begin{figure}
		\centering
		\includegraphics[width=0.8\textwidth]{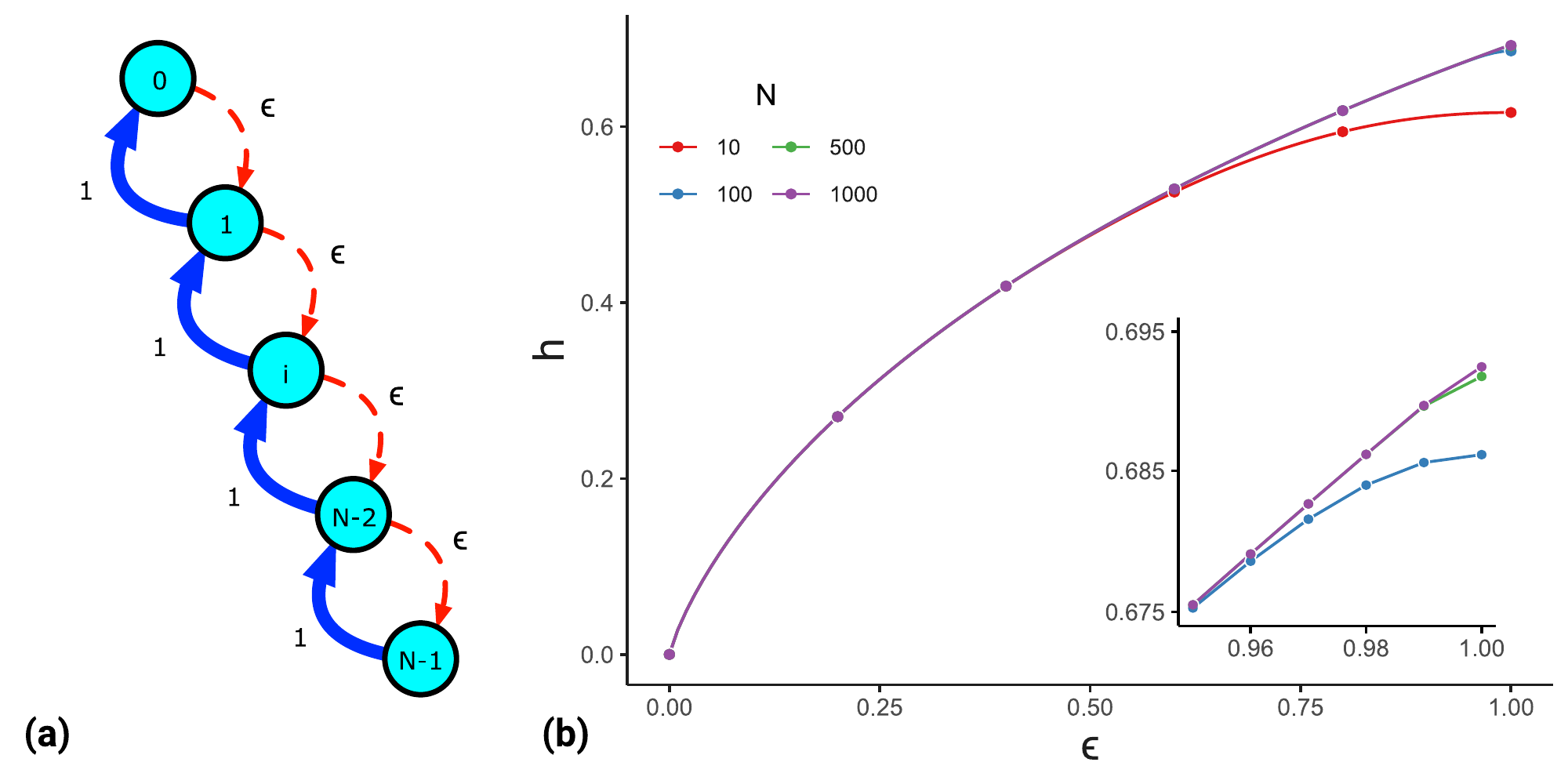}
		\caption{\textbf{Entropy rate of the generic random-walk on synthetic chain networks.} \textbf{(a)} Schematic representing the chain network {of length} $N$. The DAG structure is represented by the {blue} lines of {unitary} weight while the normality is introduced through the {red} backward edges {of} weight $\epsilon$. \textbf{(b)} entropy rate for the generic random walk taking place on these chain networks as a function of $\epsilon$ for a number of network sizes, {where} dots represent the simulated values and lines the theory as per Eq.~\eqref{eq:chain_entropy}. The three larger networks are practically indistinguishable until being close to symmetric as shown in the inset plot.}
		\label{fig:chain_model}
	\end{figure}
	
	From here it is immediately possible to determine the behavior of the extreme values of the {control parameter} $\epsilon$. First, we see that for $\lim_{\epsilon\to0} h = 0$, i.e., as the network approaches a complete hierarchical or DAG structure the entropy becomes zero as expected due to mass all accumulating with the top node in the network. The other extreme {is when the chain }network becomes entirely symmetric, and so $\lim_{\epsilon\to1} h = \dfrac{N-2}{N-1}\log(2)$. Results from simulation of {this} dynamical processes on synthetic networks alongside the corresponding estimates from Eq.~\eqref{eq:chain_entropy} are {shown} in~\ref{fig:chain_model}(b). Since this model is exactly solvable, the perfect agreement {observed} between theory and simulation {is fully expected}. 
	
	This model proves attractive as {due to} its analytical tractability allowing an insight into the {monotonic relationship that exists between the} non-normality {and} the entropy rate of the random walk. 
	
	\subsection{Non-normal Scale-free Networks}
	
	Our focus now turns to the case of a synthetic {model of non-normal} scale-free networks which are {similar to those introduced in Ref.~\cite{asllani2018structure}. With motivation coming from an extension of the original Price's model~\cite{price1965networks}, we start by first generating a scale-free} network via the configuration model~\cite{newman2010networks} which provides an undirected graph such that {the resulting} {network has} a power-law degree distribution $P(k) = k^{-\gamma}$. {Being symmetric, this network is} structurally normal {by definition. So i}n order to introduce a level of non-normality, we modify the network such that the new adjacency matrix $\tilde{A}$ is given by
	\begin{equation}
		\tilde{A} = A^{\text{upper}} + \epsilon \, A^{\text{lower}}
	\end{equation}	
	where $A^{\text{upper}}$ describes the upper triangular elements of the original adjacency matrix and likewise $A^{\text{lower}}$ describes the lower triangular elements. Note {that} this adjacency matrix reverts to the original {one} in the case $\epsilon = 1$ and in the case of $\epsilon = 0$ represents a {perfect} DAG. 
	{Although} an approximate formulation of the entropy rate {can} be found {for the case of} symmetric networks \cite{gomez2008entropy}, the asymmetric case {considered} here is {not} amenable to analysis and as such we {address the problem of calculating} the entropy rate {numerically}, \ref{fig:nnSF} shows the results of the entropy rate as a function of the strength of backward edges {and} for various values of the parameter $\gamma$. {The} simulations {are} averaged over $100$ ensembles of these networks {with a} size {of} $N = 100$ {nodes}. 
	\begin{figure}
		\centering
		\includegraphics[width=0.6\textwidth]{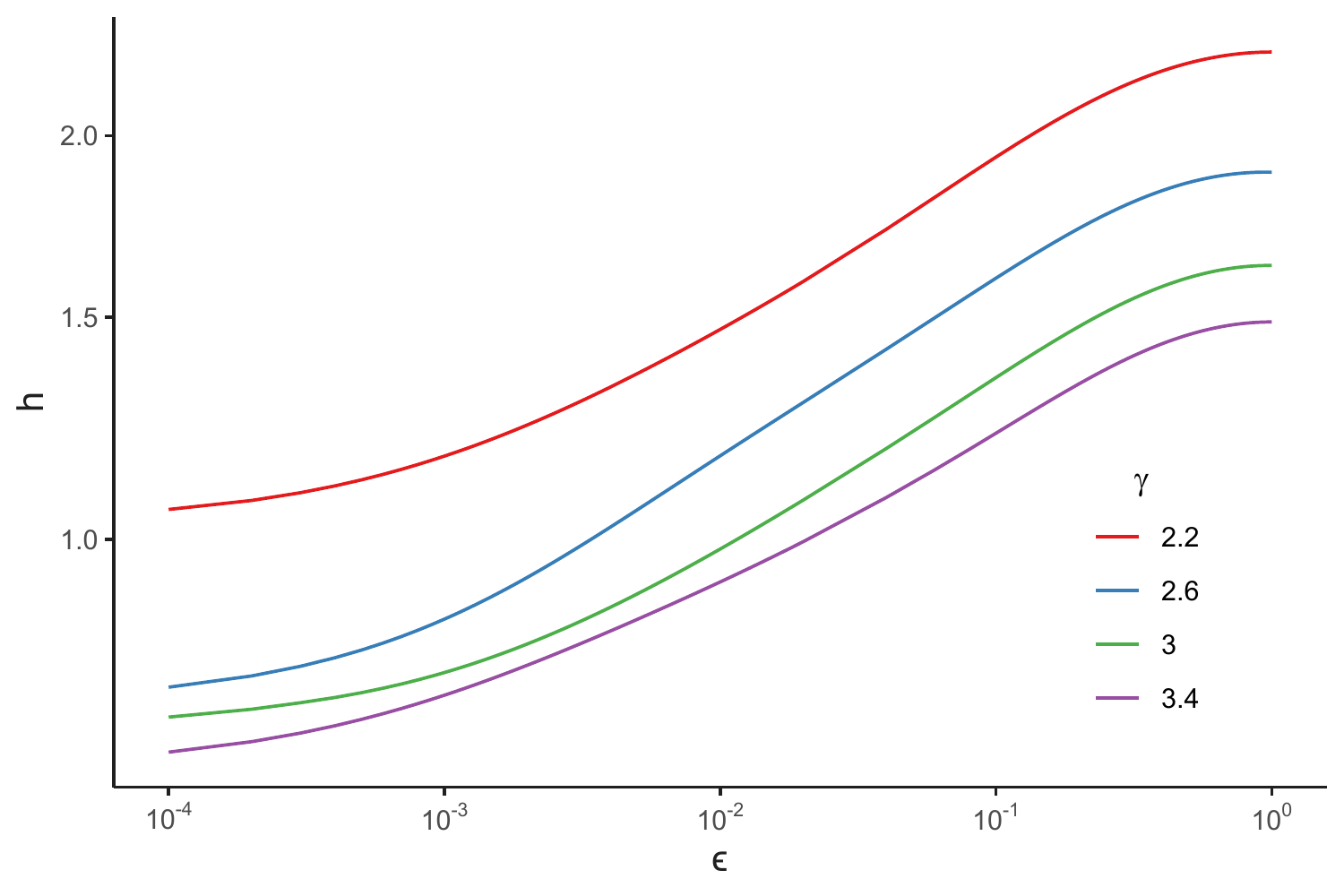}
		\caption{\textbf{Entropy rate of the generic random-walk on synthetic scale-free networks.} We generate {Non-normal} Scale-Free networks {with} backward links of weight $\epsilon$ which allows the non-normality of the network to be tuned, shown is the corresponding ER as a function of $\epsilon$ averaged over $100$ realization of networks with size $N = 100$ where the each line represents a different exponent in the degree distribution}
		\label{fig:nnSF}
	\end{figure}
	
	As in the case of the chain network, it is shown that the entropy {de}creases {while the a}symmetry (and consequently the non-normality) increases {with smaller backward edges}. This is in agreement with the {principle explained earlier} that {a balanced unstructured (random)} interaction between nodes of a {network} is positively correlated with the randomness of the process taking place upon the system. This is {also} further evidenced by the increased entropy for {networks with} degree-distributions {having} smaller exponent which results in smaller degrees for the hub nodes thus {yielding less} likelihood {of} the mass accumulating upon them.
	
	\section{Network Datasets}\label{sec:data}
	
	While we devoted our efforts into referring to the original authors responsible for collecting and first describing each of the networks, one may find some sources pointing to webpages that are long expired. In those cases, many datasets can be retrieved from repositories like the \textit{Colorado Index of Complex Networks (ICON)} at \url{https://icon.colorado.edu} or the \textit{Netzschleuder} at \url{https://networks.skewed.de}. We also highlight that the numbers corresponding to nodes, edges and so on reflect our analysis of the largest weakly connected component in each case. The quantity $F$ stands for the fraction of leader nodes out of the total described in the column Nodes (therefore a value 0 means the network has no leader nodes).
	
\begin{table}[!ht]
	\centering
	\begin{adjustbox}{max width=\textwidth}
		\begin{tabular}{llccccccc}
			\toprule
			Name & Ref. & Nodes & Edges & $\hat{d}_f$ & $\Delta$ & $h$ & $\hat{h}$ & $F$ \\
			\midrule
			\multicolumn{9}{l}{\textbf{Biological}}\\
			\cmidrule(r){1-1}
			Gene regulatory network for & \multirow{2}{*}{\cite{ref55}} & \multirow{2}{*}{328} & \multirow{2}{*}{456} &   \multirow{2}{*}{1} &   \multirow{2}{*}{1} &   \multirow{2}{*}{0} &   \multirow{2}{*}{0} & \multirow{2}{*}{0.14} \\
			\multicolumn{1}{c}{\textit{Escherichia coli} from Thieffry et al.} & & & & & & & \\ 
			Human gene regulatory network for a healthy person & \cite{ref53} & 4071 & 8466 &   1 & 0.85 & $3.5 \times 10^{-13}$ & $1.1 \times 10^{-13}$ & 0.98 \\ 
			Gene regulatory network for \textit{Saccharomyces} & \multirow{2}{*}{\cite{ref59}} & \multirow{2}{*}{2928} & \multirow{2}{*}{6149} &   \multirow{2}{*}{1} & \multirow{2}{*}{0.94} & \multirow{2}{*}{$7 \times 10^{-12}$} & \multirow{2}{*}{$1.5 \times 10^{-11}$} & \multirow{2}{*}{0.94} \\
			\multicolumn{1}{c}{\textit{cerevisiae} from Harbison et al.} & & & & & & & \\   
			Human gene regulatory network for a person with cancer & \cite{ref53} & 4049 & 11707 &   1 & 0.87 & $1.9 \times 10^{-11}$ & $5.4 \times 10^{-12}$ & 0.98 \\ 
			Gene regulatory network for \textit{Saccharomyces} & \multirow{2}{*}{\cite{ref62}} & \multirow{2}{*}{662} & \multirow{2}{*}{1063} &   \multirow{2}{*}{1} & \multirow{2}{*}{0.99} & \multirow{2}{*}{$5.6 \times 10^{-11}$} & \multirow{2}{*}{$3.4 \times 10^{-11}$} & \multirow{2}{*}{0.82} \\ 
			\multicolumn{1}{c}{\textit{cerevisiae} from Constanzo et al.} & & & & & & & \\  
			Gene regulatory network for \textit{Mycobacterium tuberculosis} & \cite{ref57} & 1604 & 3154 &   1 & 0.97 &   0 &   0 & 0.022 \\ 
			Gene regulatory network for \textit{Escherichia coli} from Salgado et al. & \cite{ref56} & 1395 & 2853 &   1 & 0.98 & $8.3 \times 10^{-11}$ & $3.2 \times 10^{-11}$ & 0.047 \\ 
			Human protein-protein interactome produced & \multirow{2}{*}{\cite{ref39}} & \multirow{2}{*}{2217} & \multirow{2}{*}{6438} &   \multirow{2}{*}{1} & \multirow{2}{*}{0.98} &   \multirow{2}{*}{0} & \multirow{2}{*}{$8.5 \times 10^{-12}$} & \multirow{2}{*}{0.089} \\
			\multicolumn{1}{c}{by a mass spectrometry‐based approach by Figeys et al.}  & & & & & & & \\ 
			Gene regulatory network for \textit{Pseudomonas aeruginosa} & \cite{ref58} & 648 & 959 & 0.99 & 0.97 &   0 &   0 & 0.051 \\ 
			Connectome of the Rhesus brain,& \multirow{2}{*}{\cite{ref70}} &  \multirow{2}{*}{91} & \multirow{2}{*}{628} & \multirow{2}{*}{0.92} & \multirow{2}{*}{0.85} & \multirow{2}{*}{$3.3 \times 10^{-10}$} & \multirow{2}{*}{0.69} & \multirow{2}{*}{0.88} \\
			\multicolumn{1}{c}{via a retrograde tracer study} & & & & & & & \\ 
			Neuronal network for \textit{Caenorhabditis elegans} & \cite{ref54} & 297 & 2345 & 0.84 & 0.73 &   0 & $1.4 \times 10^{-10}$ & 0.091 \\ 
			Neural network for \textit{Caenorhabditis elegans} frontal ganglia & \cite{ref7} & 131 & 764 & 0.84 & 0.73 & $8.1 \times 10^{-9}$ & $3.3 \times 10^{-9}$ & 0.053 \\ 
			Mouse's primary visual cortex connectome 1 & \cite{ref66} & 503 & 27667 & 0.79 & 0.64 & 0.026 & 0.8 & 0.002 \\ 
			Mouse's primary visual cortex connectome 2 & \cite{ref67} & 502 & 30079 & 0.77 & 0.61 & 3.9 & 0.82 &   0 \\ 
			Mouse's primary visual cortex connectome 3 & \cite{ref68} & 493 & 33011 & 0.7 & 0.52 & 4.1 & 0.86 &   0 \\ 
			Connectome of the Rhesus brain, extracted from tract & \multirow{2}{*}{\cite{ref69}} & \multirow{2}{*}{242} & \multirow{2}{*}{4090} & \multirow{2}{*}{0.65} & \multirow{2}{*}{0.41} & \multirow{2}{*}{2.9} & \multirow{2}{*}{0.85} &   \multirow{2}{*}{0} \\ 
			\multicolumn{1}{c}{tracing studies collated in the CoCoMac database} & & & & & & & \\
			Metabolic network for \textit{Saccharomyces cerevisiae} & \cite{ref52} & 1510 & 3833 & 0.64 & 0.47 & $8.8 \times 10^{-219}$ & $1.9 \times 10^{-8}$ & 0.026 \\ 
			Neuronal network for a mouse brain & \cite{ref28} & 213 & 21654 & 0.59 & 0.28 & 4.5 & 0.91 &   0 \\ 
			Connectome of a cat & \cite{ref43} &  65 & 1139 & 0.49 & 0.24 & 2.8 & 0.89 &   0 \\ 
			\midrule
			\multicolumn{9}{l}{\textbf{Ecological}}\\
			\cmidrule(r){1-1}  
			River Foodweb in Berwick Stream, New Zealand & \cite{ref4} &  77 & 240 &   1 &   1 &   0 &   0 & 0.45 \\ 
			River Foodweb in Black Rock Stream, New Zealand & \cite{ref5} &  86 & 375 &   1 & 0.97 &   0 &   0 & 0.57 \\ 
			River Foodweb in Broad Stream, New Zealand & \cite{ref5} &  94 & 564 &   1 & 0.98 &   0 &   0 & 0.56 \\ 
			River Foodweb in Caitlins Stream, New Zealand & \cite{ref9} &  48 & 110 &   1 &   1 &   0 &   0 & 0.29 \\ 
			River Foodweb in Coweeta, USA & \cite{ref4} &  71 & 148 &   1 &   1 &   0 &   0 & 0.54 \\ 
			River Foodweb in Coweeta, USA & \cite{ref4} &  58 & 126 &   1 &   1 &   0 &   0 & 0.48 \\ 
			River Foodweb in Dempsters Stream during autumn, New Zealand & \cite{ref9} &  83 & 414 &   1 & 0.92 &   0 &   0 & 0.55 \\ 
			River Foodweb in Dempsters Stream during spring, New Zealand & \cite{ref9} &  93 & 538 &   1 & 0.97 &   0 &   0 & 0.54 \\ 
			River Foodweb in Dempsters Stream during summer, New Zealand & \cite{ref5} & 107 & 965 &   1 & 0.95 &   0 &   0 & 0.47 \\ 
			Marine Foodweb in Flensburg Fjord, Germany/Denmark & \cite{ref2} &  77 & 576 &   1 & 0.91 &   0 &   0 & 0.078 \\ 
			River Foodweb in German Creek, New Zealand & \cite{ref5} &  84 & 352 &   1 & 0.99 &   0 &   0 & 0.57 \\ 
			River Foodweb in Healy Creek, New Zealand & \cite{ref5} &  96 & 634 &   1 &   1 &   0 &   0 & 0.49 \\ 
			River Foodweb in Kye Burn, New Zealand & \cite{ref5} &  98 & 629 &   1 & 0.89 &   0 &   0 & 0.59 \\ 
			River Foodweb in Little Kye Burn, New Zealand & \cite{ref5} &  78 & 375 &   1 &   1 &   0 &   0 & 0.54 \\ 
			River Foodweb in Martins Stream, USA & \cite{ref4} & 105 & 343 &   1 & 0.98 &   0 &   0 & 0.46 \\ 
			River Foodweb in Narrowdale Stream, New Zealand & \cite{ref13} &  71 & 154 &   1 &   1 &   0 &   0 & 0.39 \\ 
			River Foodweb in North Col Stream, New Zealand & \cite{ref13} &  78 & 241 &   1 &   1 &   0 &   0 & 0.32 \\ 
			River Foodweb in Powder Stream, New Zealand & \cite{ref4} &  78 & 268 &   1 & 0.95 &   0 &   0 & 0.41 \\ 
			River Foodweb in Stony Stream, New Zealand & \cite{ref5} & 112 & 830 &   1 & 0.97 &   0 &   0 & 0.56 \\ 
			River Foodweb in Sutton Stream during autumn, New Zealand & \cite{ref9} &  80 & 335 &   1 &   1 &   0 &   0 & 0.61 \\ 
			River Foodweb in Sutton Stream during spring, New Zealand & \cite{ref9} &  74 & 391 &   1 & 0.92 &   0 &   0 & 0.68 \\ 
			River Foodweb in Sutton Stream during summer, New Zealand & \cite{ref5} &  87 & 424 &   1 &   1 &   0 &   0 & 0.72 \\ 
			River Foodweb in Troy Stream, USA & \cite{ref4} &  77 & 181 &   1 & 0.91 &   0 &   0 & 0.52 \\ 
			River Foodweb in Venlaw Stream, New Zealand & \cite{ref4} &  66 & 187 &   1 &   1 &   0 &   0 & 0.45 \\ 
			Terrestrial Foodweb in Scotch broom, England & \cite{ref22} &  85 & 219 &   1 &   1 &   0 &   0 & 0.6 \\ 
			River Foodweb in Canton Creek, New Zealand & \cite{ref5} & 102 & 696 &   1 & 0.98 &   0 &   0 & 0.53 \\ 
			Marine Foodweb in Cayman Islands & \cite{ref23} & 242 & 3764 &   1 & 0.99 & $5.4 \times 10^{-16}$ & $1.5 \times 10^{-16}$ & 0.041 \\ 
			Marine Foodweb in Chesapeake Bay, USA & \cite{ref24} &  31 &  67 &   1 & 0.85 &   0 &   0 & 0.16 \\ 
			\bottomrule
		\end{tabular}
	\end{adjustbox}
\end{table}

\begin{table}[!ht]
	\centering
	\begin{adjustbox}{max width=\textwidth}
		\begin{tabular}{llccccccc}
			\toprule
			Name & Ref. & Nodes & Edges & $\hat{d}_f$ & $\Delta$ & $h$ & $\hat{h}$ & $F$ \\ 
			\midrule
			\multicolumn{9}{l}{\textbf{Ecological}}\\
			\cmidrule(r){1-1}  
			Terrestrial Foodweb in grasslands of the United Kingdom & \cite{ref28} &  61 &  97 &   1 & 0.98 &   0 &   0 & 0.13 \\ 
			Dominance among ants & \cite{ref44} &  16 &  36 &   1 & 0.13 &   0 &   0 & 0.12 \\ 
			Dominance among kangaroos & \cite{ref47} &  17 &  91 &   1 & 0.97 &   0 &   0 & 0.24 \\ 
			Marine Foodweb in St. Marks Estuary, US & \cite{ref75} &  48 & 218 &   1 & 0.85 &   0 &   0 & 0.12 \\ 
			Terrestrial Foodweb in Saint-Martin Island, Lesser Antilles & \cite{ref76} &  42 & 205 &   1 &   1 &   0 &   0 & 0.14 \\ 
			River Foodweb in Stony Stream, New Zealand & \cite{ref5} & 109 & 827 &   1 &   1 &   0 &   0 & 0.56 \\ 
			Marine Foodweb in Ythan Estuary, Scotland & \cite{ref19} &  82 & 391 &   1 & 0.99 & $3.3 \times 10^{-12}$ & $1.6 \times 10^{-12}$ & 0.37 \\ 
			Lake Foodweb in Lough Hyne, Ireland & \cite{ref12} & 349 & 5102 &   1 & 0.98 & $2 \times 10^{-11}$ & $5.5 \times 10^{-12}$ & 0.14 \\ 
			Fossil Assemblage Foodweb from Burgess Shale, Canada & \cite{ref6} &  48 & 243 & 0.99 & 0.99 & $4 \times 10^{-13}$ & $1.6 \times 10^{-13}$ & 0.12 \\ 
			Lake Foodweb in Bridge Broom Lake & \cite{ref21} &  25 & 104 & 0.99 & 0.96 & $1.3 \times 10^{-12}$ & $2 \times 10^{-11}$ & 0.04 \\ 
			Dominance among wolves & \cite{ref51} &  16 & 148 & 0.98 & 0.92 & $1.3 \times 10^{-10}$ & 0.7 & 0.062 \\ 
			Lake Foodweb in Little Rock Lake, USA & \cite{ref36} & 183 & 2476 & 0.98 & 0.93 & $3.1 \times 10^{-12}$ & 0.36 & 0.0055 \\ 
			Marine Foodweb in Northeast United States Shelf & \cite{ref73} &  79 & 1378 & 0.98 & 0.99 & $1.2 \times 10^{-11}$ & $3.4 \times 10^{-12}$ & 0.025 \\ 
			Aggression expressed by juvenile & \multirow{2}{*}{\cite{ref41}} &  \multirow{2}{*}{16} &  \multirow{2}{*}{97} & \multirow{2}{*}{0.98} & \multirow{2}{*}{0.89} & \multirow{2}{*}{$8.4 \times 10^{-12}$} & \multirow{2}{*}{$3.8 \times 10^{-12}$} & \multirow{2}{*}{0.062} \\ 
			\multicolumn{1}{c}{\textit{Macaca nemestrina} towards others} & & & & & & & \\
			Competition for an object or to occupy & \multirow{2}{*}{\cite{ref41}} &  \multirow{2}{*}{16} &  \multirow{2}{*}{97} & \multirow{2}{*}{0.98} & \multirow{2}{*}{0.74} & \multirow{2}{*}{$8.4 \times 10^{-12}$} & \multirow{2}{*}{$3.8 \times 10^{-12}$} & \multirow{2}{*}{0.062} \\ 
			\multicolumn{1}{c}{space among juvenile \textit{Macaca nemestrina}} & & & & & & & \\
			Lake Foodweb in Skipwith Common, England & \cite{ref74} &  25 & 189 & 0.98 & 0.96 & $5.6 \times 10^{-13}$ & $2 \times 10^{-13}$ & 0.04 \\ 
			Marine Foodweb in Weddel Sea, Antarctica & \cite{ref12} & 483 & 15317 & 0.98 & 0.95 & $3.8 \times 10^{-10}$ & $8.6 \times 10^{-11}$ & 0.13 \\ 
			Marine Foodweb in Benguela Current, South Africa & \cite{ref20} &  29 & 196 & 0.97 & 0.94 & $3.6 \times 10^{-12}$ & $1.4 \times 10^{-12}$ & 0.069 \\ 
			Marine Foodweb in Florida Bay during dry season & \cite{ref32} & 128 & 2137 & 0.97 & 0.84 & $4.1 \times 10^{-11}$ & $3.5 \times 10^{-11}$ & 0.0078 \\ 
			Dominance among macaques & \cite{ref48} &  62 & 1187 & 0.97 & 0.93 & $2.7 \times 10^{-11}$ & $1.4 \times 10^{-11}$ & 0.016 \\ 
			Terrestrial Foodweb in El Verde Field Station, Puerto Rico & \cite{ref28} & 155 & 1507 & 0.94 & 0.85 & $1.6 \times 10^{-9}$ & $4.8 \times 10^{-10}$ & 0.18 \\ 
			Dominance among ponies & \cite{ref49} &  17 & 146 & 0.93 & 0.74 & $1.4 \times 10^{-10}$ & $5.6 \times 10^{-11}$ & 0.059 \\ 
			Dominance among cattle & \cite{ref46} &  28 & 217 & 0.92 & 0.6 & $5.3 \times 10^{-11}$ & $3.3 \times 10^{-8}$ & 0.036 \\ 
			Dominance among sheep & \cite{ref50} &  28 & 250 & 0.89 & 0.78 & $1.9 \times 10^{-11}$ & $7 \times 10^{-12}$ & 0.036 \\ 
			Fossil Assemblage Foodweb from Chengjiang Shale, China & \cite{ref6} &  33 &  90 & 0.86 & 0.71 & $4.1 \times 10^{-10}$ & $2.2 \times 10^{-10}$ & 0.12 \\ 
			Dominance among white leghorn hens & \cite{ref77} &  32 & 496 & 0.86 & 0.74 & $4.5 \times 10^{-10}$ & $1.3 \times 10^{-10}$ & 0.031 \\ 
			Marine Foodweb in Estero de Punta Banda, Mexico & \cite{ref2} & 143 & 3696 & 0.84 & 0.73 & $3.5 \times 10^{-9}$ & $9 \times 10^{-10}$ & 0.035 \\ 
			Dominance among bisons & \cite{ref45} &  26 & 314 & 0.76 & 0.55 & 2.1 & 0.83 &   0 \\ 
			Marine Foodweb in Carpinteria Salt Marsh Reserve, USA & \cite{ref2} & 166 & 7682 & 0.72 & 0.53 & $1.6 \times 10^{-8}$ & $3.6 \times 10^{-9}$ & 0.03 \\ 
			Marine Foodweb in Ythan Estuary, Scotland & \cite{ref2} & 166 & 9029 & 0.69 & 0.48 & $2.3 \times 10^{-8}$ & $5.2 \times 10^{-9}$ & 0.03 \\ 
			Marine Foodweb in Bahia Falsa, Mexico & \cite{ref2} & 166 & 9576 & 0.68 & 0.47 & $2.6 \times 10^{-8}$ & $5.8 \times 10^{-9}$ & 0.03 \\ 
			Marine Foodweb in Sylt Tidal Basin, Germany & \cite{ref2} & 215 & 14963 & 0.66 & 0.44 & $2.4 \times 10^{-6}$ & $5 \times 10^{-7}$ & 0.0047 \\ 
			Marine Foodweb in Otago Harbour, New Zealand & \cite{ref2} & 215 & 15266 & 0.65 & 0.41 & $2.4 \times 10^{-6}$ & $5 \times 10^{-7}$ & 0.0047 \\ 
			Affiliative interaction network between giraffes & \cite{ref33} &   6 &  30 & 0.34 & 0.11 & 1.5 & 0.98 &   0 \\ 
			Nearest neighbour network between giraffes & \cite{ref33} &   6 &  30 & 0.15 & 0.03 & 1.6 &   1 &   0 \\ 
			Proximity network between giraffes & \cite{ref33} &   6 &  30 & 0.13 & 0.029 & 1.6 &   1 &   0 \\ 
			\midrule
			\multicolumn{9}{l}{\textbf{Economic}}\\
			\cmidrule(r){1-1}
			Finance flows between organizations in & \multirow{2}{*}{\cite{ref42}} &  \multirow{2}{*}{61} &  \multirow{2}{*}{60} &   \multirow{2}{*}{1} & \multirow{2}{*}{0.93} &   \multirow{2}{*}{0} &   \multirow{2}{*}{0} & \multirow{2}{*}{0.74} \\
			\multicolumn{1}{c}{the Bernie Madoff ponzi scheme} & & & & & & & \\  
			International currencies network in 1890 & \cite{ref31} &  45 & 194 & 0.77 & 0.6 & $3.9 \times 10^{-9}$ & 0.82 & 0.51 \\ 
			International currencies network in 1900 & \cite{ref31} &  45 & 218 & 0.74 & 0.56 & $5.6 \times 10^{-10}$ & 0.83 & 0.44 \\ 
			International currencies network in 1910 & \cite{ref31} &  45 & 264 & 0.72 & 0.53 &   0 & 0.86 & 0.4 \\ 
			Aggregated multiplex trade network from FAO & \cite{ref30} & 214 & 13715 & 0.65 & 0.41 & $2.3 \times 10^{-8}$ & 0.88 & 0.38 \\ 
			World trade network & \cite{ref60} &  78 & 923 & 0.62 & 0.54 & $2.1 \times 10^{-9}$ & $8.7 \times 10^{-10}$ & 0.28 \\ 
			International trade network of minerals & \cite{ref61} &  24 & 135 & 0.61 & 0.39 & $4 \times 10^{-9}$ & $1.8 \times 10^{-9}$ & 0.12 \\ 
			International trade network of manufactured food products & \cite{ref61} &  24 & 307 & 0.49 & 0.28 & 2.4 & 0.86 &   0 \\ 
			International trade network of manufactured goods & \cite{ref61} &  24 & 310 & 0.48 & 0.24 & $7.1 \times 10^{-9}$ & $2.5 \times 10^{-9}$ & 0.083 \\ 
			International trade network of crude animal & \multirow{2}{*}{\cite{ref61}} &  \multirow{2}{*}{24} & \multirow{2}{*}{307} & \multirow{2}{*}{0.46} & \multirow{2}{*}{0.24} & \multirow{2}{*}{2.5} & \multirow{2}{*}{0.91} &   \multirow{2}{*}{0} \\
			\multicolumn{1}{c}{and vegetable material} & & & & & & & \\ 
			International trade network of diplomatic exchanges & \cite{ref61} &  24 & 369 & 0.27 & 0.079 & 2.8 & 0.97 &   0 \\
			\bottomrule
		\end{tabular}
	\end{adjustbox}
\end{table}

\begin{table}[!ht]
	\centering
	\begin{adjustbox}{max width=\textwidth}
		\begin{tabular}{llccccccc}
			\toprule
			Name & Ref. & Nodes & Edges & $\hat{d}_f$ & $\Delta$ & $h$ & $\hat{h}$ & $F$ \\ 
			\midrule
			\multicolumn{9}{l}{\textbf{Informational}}\\
			\cmidrule(r){1-1}
			Citations from papers that cite "Small World Problem” & \cite{ref17} & 233 & 994 &   1 &   1 &   0 & $8.9 \times 10^{-14}$ & 0.0086 \\ 
			Citations to Small, Griffith and descendants & \cite{ref16} & 1024 & 4918 &   1 & 0.96 & $5 \times 10^{-13}$ & $4 \times 10^{-13}$ & 0.02 \\ 
			Citation network of the journal Scientometrics & \cite{ref14} & 2678 & 10381 &   1 & 0.91 &   0 &   0 & 0.34 \\ 
			Papers citing and by AH Zewail & \cite{ref17} & 6651 & 54232 &   1 & 0.87 &   0 &   0 & 0.31 \\ 
			Citations among papers contained in the DBLP & \cite{ref26} & 12591 & 49728 &   1 & 0.64 &   0 & $3.7 \times 10^{-7}$ & 0.082 \\ 
			\multicolumn{1}{c}{computer science bibliography as of May of 2014}  & & & & & & & \\
			Word adjacency network for a japanese book &  \multirow{2}{*}{\cite{ref34}} &  \multirow{2}{*}{2698} &  \multirow{2}{*}{8297} &  \multirow{2}{*}{0.9} &  \multirow{2}{*}{0.44} &  \multirow{2}{*}{$2.5 \times 10^{-8}$} &  \multirow{2}{*}{$2 \times 10^{-8}$} &  \multirow{2}{*}{0.26} \\ 
			Word adjacency network for & \multirow{2}{*}{\cite{ref28}} &  \multirow{2}{*}{50} & \multirow{2}{*}{101} & \multirow{2}{*}{0.82} & \multirow{2}{*}{0.52} & \multirow{2}{*}{$3.8 \times 10^{-9}$} & \multirow{2}{*}{$2.2 \times 10^{-9}$} & \multirow{2}{*}{0.32} \\
			\multicolumn{1}{c}{Dr. Seuss’s Green Eggs and Ham book} & & & & & & & \\
			\midrule
			\multicolumn{9}{l}{\textbf{Social}}\\
			\cmidrule(r){1-1} 
			Literary criticism network of Dutch writers in 1976 & \cite{ref35} &  35 &  81 & 0.99 & 0.9 & $9.9 \times 10^{-12}$ & $6.5 \times 10^{-11}$ & 0.31 \\ 
			Offensive media-sharing network from Brazilian Federal Police & \cite{ref27} & 10407 & 842247 & 0.98 & 0.79 & $2.7 \times 10^{-9}$ & $4.4 \times 10^{-10}$ & 0.00029 \\ 
			Mentions network of identified Russian troll accounts & \cite{ref72} & 1149 & 2656 & 0.96 & 0.91 &   0 & $1.9 \times 10^{-175}$ & 0.73 \\ 
			Votes on admin elections of Wikipedia in 2008 & \cite{ref18} & 7115 & 103689 & 0.94 & 0.73 & $8.5 \times 10^{-9}$ & $6.6 \times 10^{-9}$ & 0.67 \\ 
			Political Blogs Network & \cite{ref1} & 1222 & 18956 & 0.84 & 0.68 &   0 &   0 & 0.13 \\ 
			Chess players and outcomes between their matches & \cite{ref25} & 7301 & 60046 & 0.8 & 0.21 & 0.00019 & 0.00046 & 0.098 \\ 
			Friendship among college students in a course about leadership & \cite{ref34} &  32 &  96 & 0.78 & 0.58 & $2 \times 10^{-9}$ & 0.68 & 0.16 \\ 
			Trust relationships among physicians in four towns in Illinois & \cite{ref8} & 117 & 543 & 0.77 & 0.62 & $7 \times 10^{-9}$ & $4.8 \times 10^{-9}$ & 0.094 \\ 
			E-mail network for the Democratic National Convention & \cite{ref10} & 1833 & 5499 & 0.75 & 0.5 &   0 & $1.9 \times 10^{-10}$ & 0.28 \\ 
			Friendship among highschool students & \cite{ref28} &  70 & 366 & 0.65 & 0.46 & $1.4 \times 10^{-8}$ & 0.83 & 0.043 \\ 
			Online messages from an online community of & \multirow{2}{*}{\cite{ref64}} & \multirow{2}{*}{1893} & \multirow{2}{*}{20292} & \multirow{2}{*}{0.59} & \multirow{2}{*}{0.3} &   \multirow{2}{*}{0} &   \multirow{2}{*}{0} & \multirow{2}{*}{0.29} \\
			\multicolumn{1}{c}{students from the University of California, Irvine} & & & & & & & \\
			\midrule
			\multicolumn{9}{l}{\textbf{Transport}}\\
			\cmidrule(r){1-1}  
			Roads Network in Winnipeg, Canada & \cite{ref3} & 949 & 1823 &   1 & 0.13 &   0 &   0 & 0.0011 \\ 
			Roads Network in Barcelona, Spain & \cite{ref3} & 930 & 2522 & 0.9 & 0.25 & $8 \times 10^{-6}$ & $6.8 \times 10^{-6}$ & 0.0011 \\ 
			Air-traffic control network from FAA & \cite{ref29} & 1226 & 2613 & 0.82 & 0.57 & $1 \times 10^{-8}$ & $6.1 \times 10^{-9}$ & 0.12 \\ 
			Roads Network in Terrassa, Spain & \cite{ref3} & 1603 & 3264 & 0.8 & 0.14 & 0.4 & 0.87 &   0 \\ 
			Roads Network in Sioux Falls, USA & \cite{ref15} &  24 &  76 & 0.46 & 0.11 &   1 &   1 &   0 \\ 
			Flights between airports in the United States & \cite{ref65} & 1572 & 28235 & 0.44 & 0.094 & $2.1 \times 10^{-121}$ & $9.1 \times 10^{-7}$ & 0.044 \\ 
			London bike-sharing network & \cite{ref37} & 750 & 187713 & 0.35 & 0.085 &   5 & 0.98 &   0 \\ 
			Roads Network in Rome, Italy & \cite{ref71} & 3353 & 8859 & 0.32 & 0.068 & 0.9 & 0.94 &   0 \\ 
			London tube network & \cite{ref38} & 270 & 628 & 0.23 & 0.002 & 0.82 &   1 &   0 \\ 
			Flights between world airports from openflights.org & \cite{ref63} & 2905 & 30442 & 0.22 & 0.021 & $5.2 \times 10^{-6}$ & $1.5 \times 10^{-6}$ & 0.0062 \\ 
			Paris metropolitan train grid & \cite{ref38} & 302 & 705 & 0.14 & 0.0089 & 0.9 & 0.98 &   0 \\ 
			Roads Network in Eastern Massachussets, USA & \cite{ref11} &  74 & 258 & 0.076 & 0.023 & 1.2 &   1 &   0 \\ 
			\bottomrule
		\end{tabular}
	\end{adjustbox}
\end{table}
	
	
	\section{Hierarchical Analysis}
	
	In the main text we comment on the hierarchical structure of the empirical networks under consideration. This analysis is conducted by constructing a labeling scheme for each network. This is completed by first first identifying all leader nodes and then proceeding to look for shortest paths from each of these nodes to each node in the leader's in-component i.e., the nodes from which it can be reached. The nodes in this in-component are given a label based upon their distance to the corresponding leader. Of course, in networks where there are more than one leader the ancillary nodes may have multiple labels and as such the last step is to take the minimum of the labels i.e., the distance to the nearest leader. A full description of this procedure is given in Algorithm \ref{algo:hierarchy}.
	
	\begin{algorithm}[H]
		\caption{Hierarchical label identification for each node in a given network of size $N$. Firstly, it finds all leaders or nodes with no outdegree. If there are no leaders the algorithm stops and does not return labels. Otherwise, it proceeds to find shortest paths over the network from each of these leaders. Every node is given a label corresponding to their minimum distance to a leader node.}
		\label{algo:hierarchy}
		\begin{algorithmic}[1]
			\State \textit{inputs:} $\mathbf{A}$
			\State \textit{outputs:} $L$
			\State $\text{deg} \leftarrow \text{outdegree}(\mathbf{A})$
			\State $\text{leaders} \leftarrow []$
			\For {$j = 1$ to $N$}
			\If{deg$(j) = 0$}
			\State leaders.append$(j)$
			\State $L[j] \leftarrow 0$
			\Else
			\State $L[j] \leftarrow N$ 
			\EndIf
			\EndFor
			\State $\ell \leftarrow$  size(leaders) 
			\If{$\ell = 0$}
			\State \Return $[]$
			\Else
			\State $\mathbf{A} \leftarrow \text{transpose}(\mathbf{A})$
			\State $\mathbf{G} \leftarrow \text{digraph}(\mathbf{A})$
			\For {$i = 1$ to $\ell$}
			\State leader $\leftarrow$ leaders$(i)$
			\State labels $\leftarrow$ shortestpathdistancesfrom($\mathbf{G}$,leader)
			\For{$j = 1$ to $N$}
			\State $L[j] \leftarrow \min(L[j], \text{labels}[j])$
			\EndFor
			\EndFor
			\State \Return $L$
			\EndIf
		\end{algorithmic}
	\end{algorithm}

	With this labels at hand we proceed to conduct analysis with regards the structure of edges within each network. In particular, we focus upon the hierarchical levels of nodes at the end of each edge in the network. We then view edges based upon their contribution towards the networks structure such that those joining a larger to smaller hierarchical level (and thus contributing towards the leader nodes) are blue, those in the opposite direction (contributing away from the leader) are red, while lastly those which join two nodes of the same hierarchical level are green. Furthermore, we also provide analysis into the  general structure by looking at the $H \times H$ matrix, where $H$ is the number of hierarchical levels, with entries $\psi_{ij}$ describing the sum of the weights of edges joining nodes in hierarchical level $j$ to those in hierarchical level $i$. Importantly, the upper triangular elements of this matrix describes the blue edges, the diagonal elements the green edges, and the lower triangular the red edges.
	
	We proceed to visualize, as in Fig.~5 of the main text, the structure of each network in our collection. For each case we provide three visualizations:
	\begin{enumerate}
		\item We consider only the blue and red edges within the network and demonstrate the fraction of these edges to (from) nodes with a given hierarchical in the case of the blue (red) edges;
		\item A heatmap representation of the $\psi_{ij}$ entries in for each $i,j$ as described above;
		\item A bar chart representation of the total weight of edges with each color where we have included blue and green together as they both can be viewed as aiding those with the same/higher hierarchical position. 
	\end{enumerate}
	
	\begin{figure}[h]
		\centering
		\includegraphics[width=\textwidth]{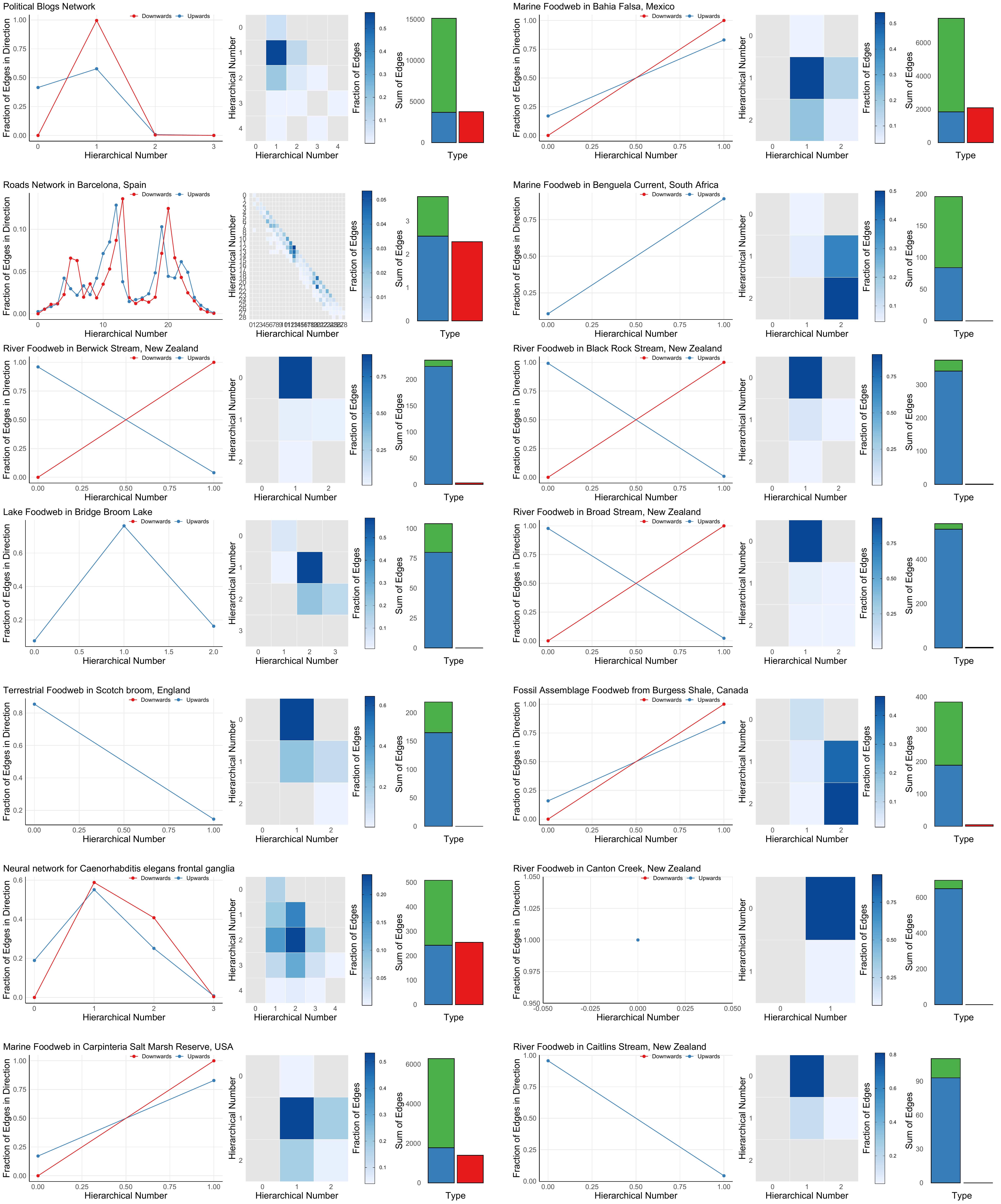}
	\end{figure}
	
	\begin{figure}
		\centering
		\includegraphics[width=\textwidth]{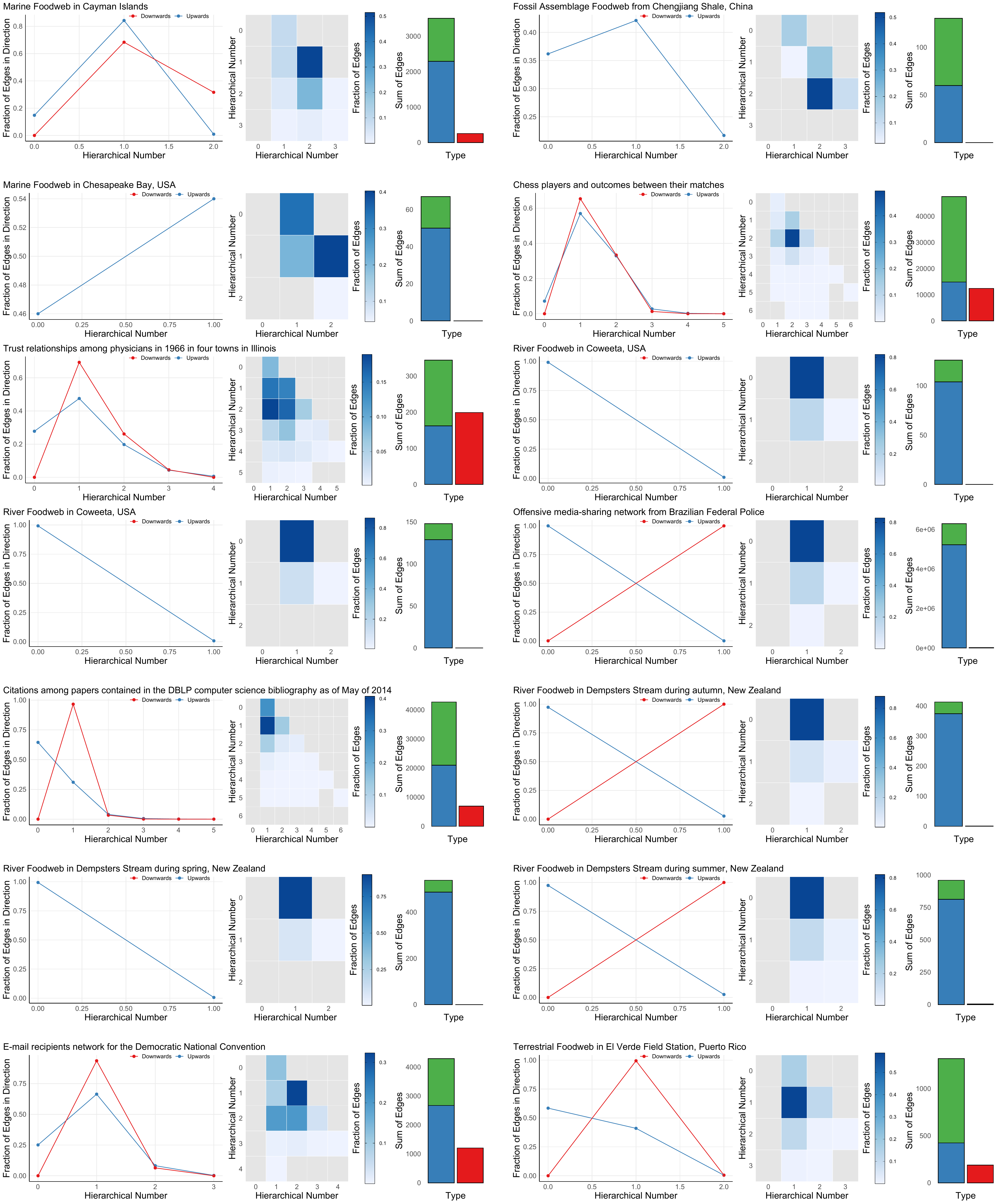}
	\end{figure}
	
	\begin{figure}
		\centering
		\includegraphics[width=\textwidth]{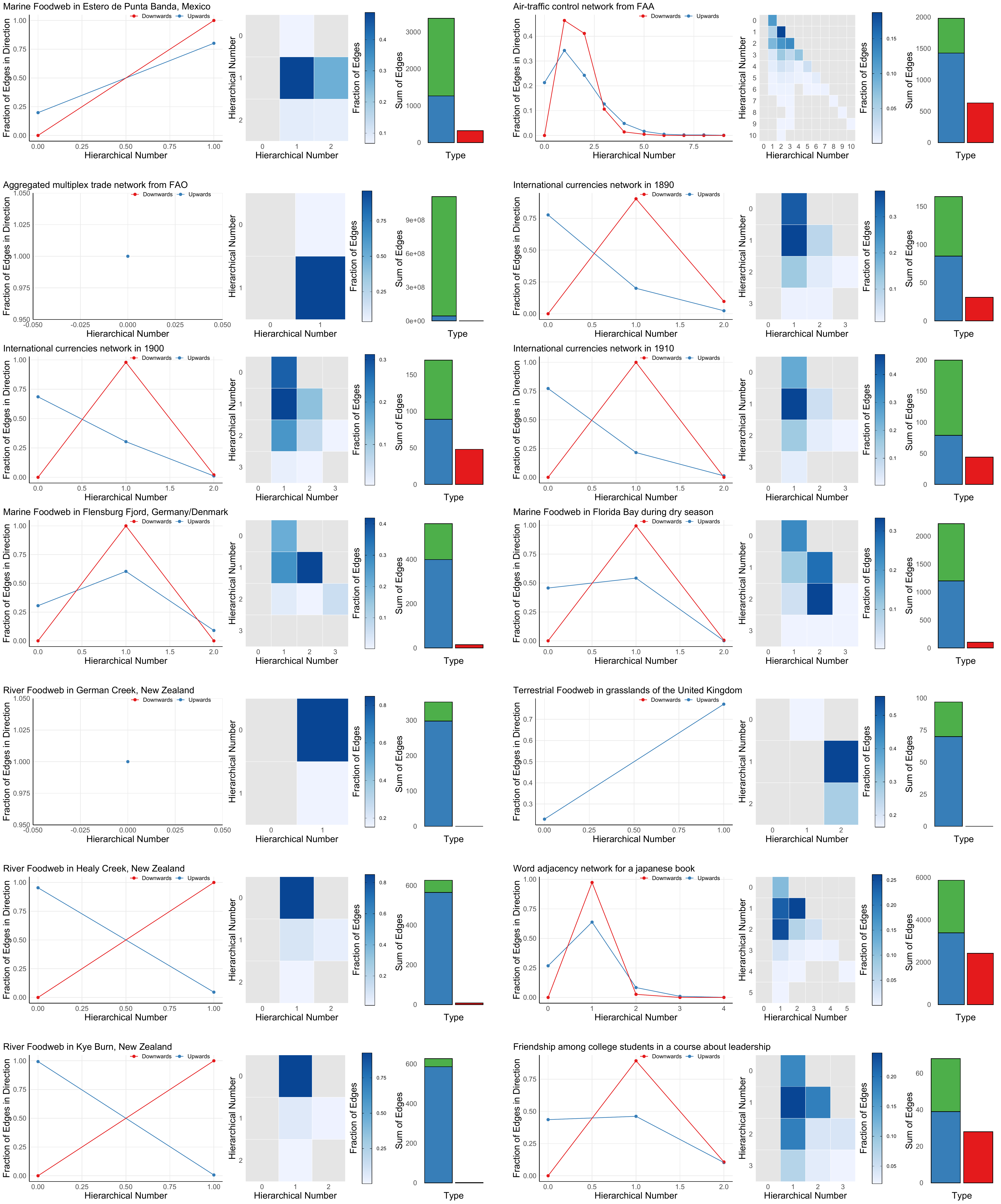}
	\end{figure}
	
	\begin{figure}
		\centering
		\includegraphics[width=\textwidth]{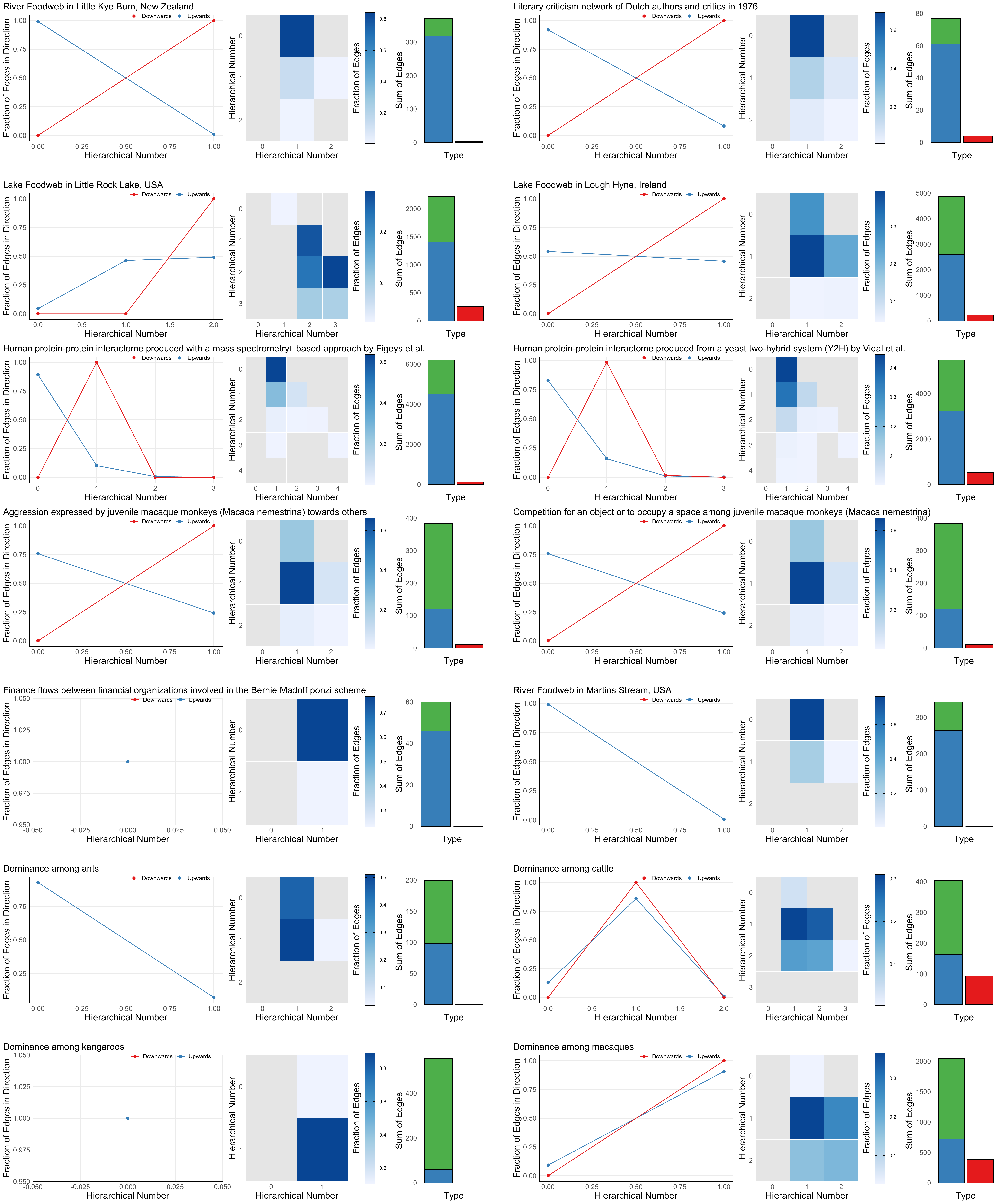}
	\end{figure}
	
	\begin{figure}
		\centering
		\includegraphics[width=\textwidth]{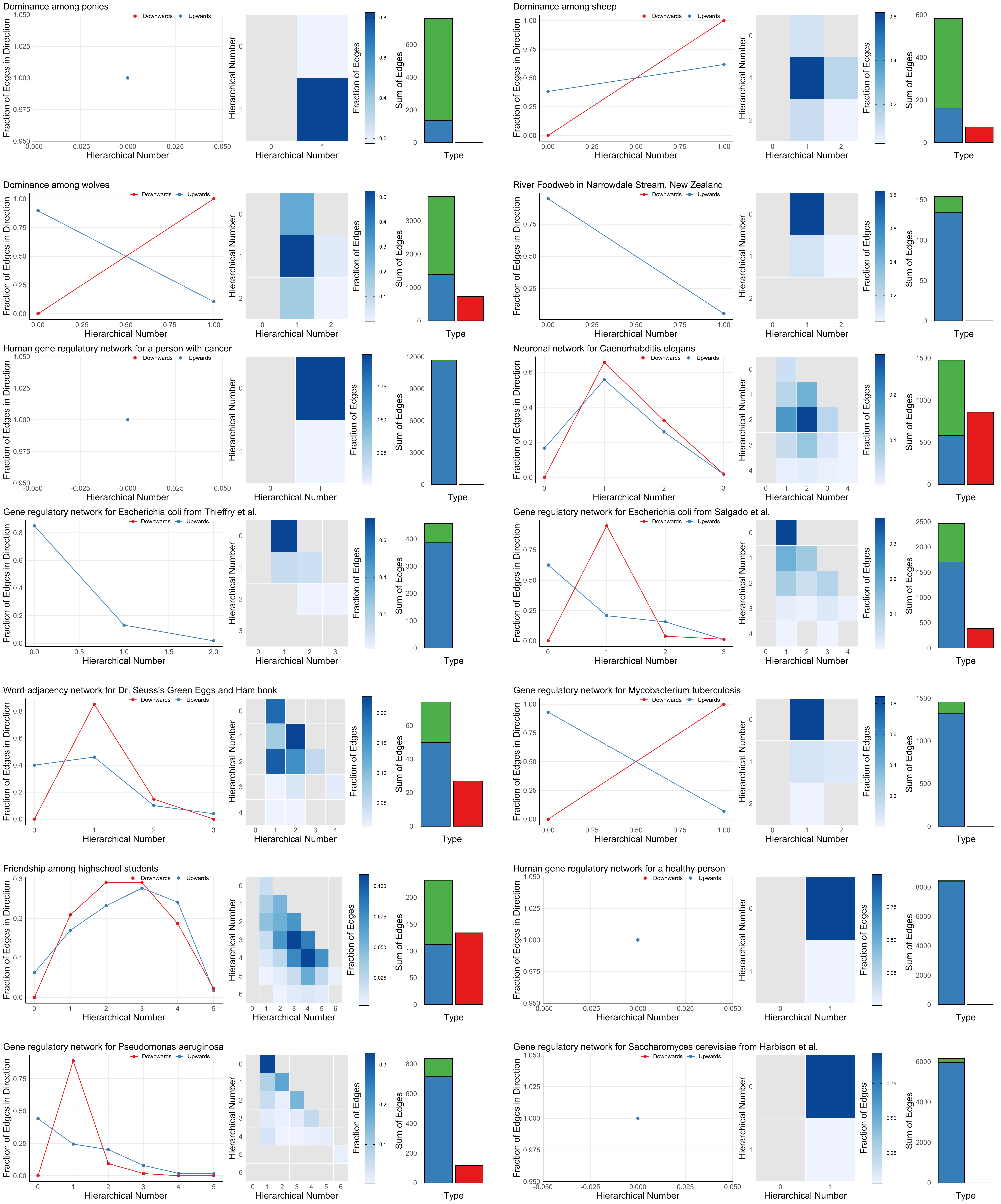}
	\end{figure}
	
	\begin{figure}
		\centering
		\includegraphics[width=\textwidth]{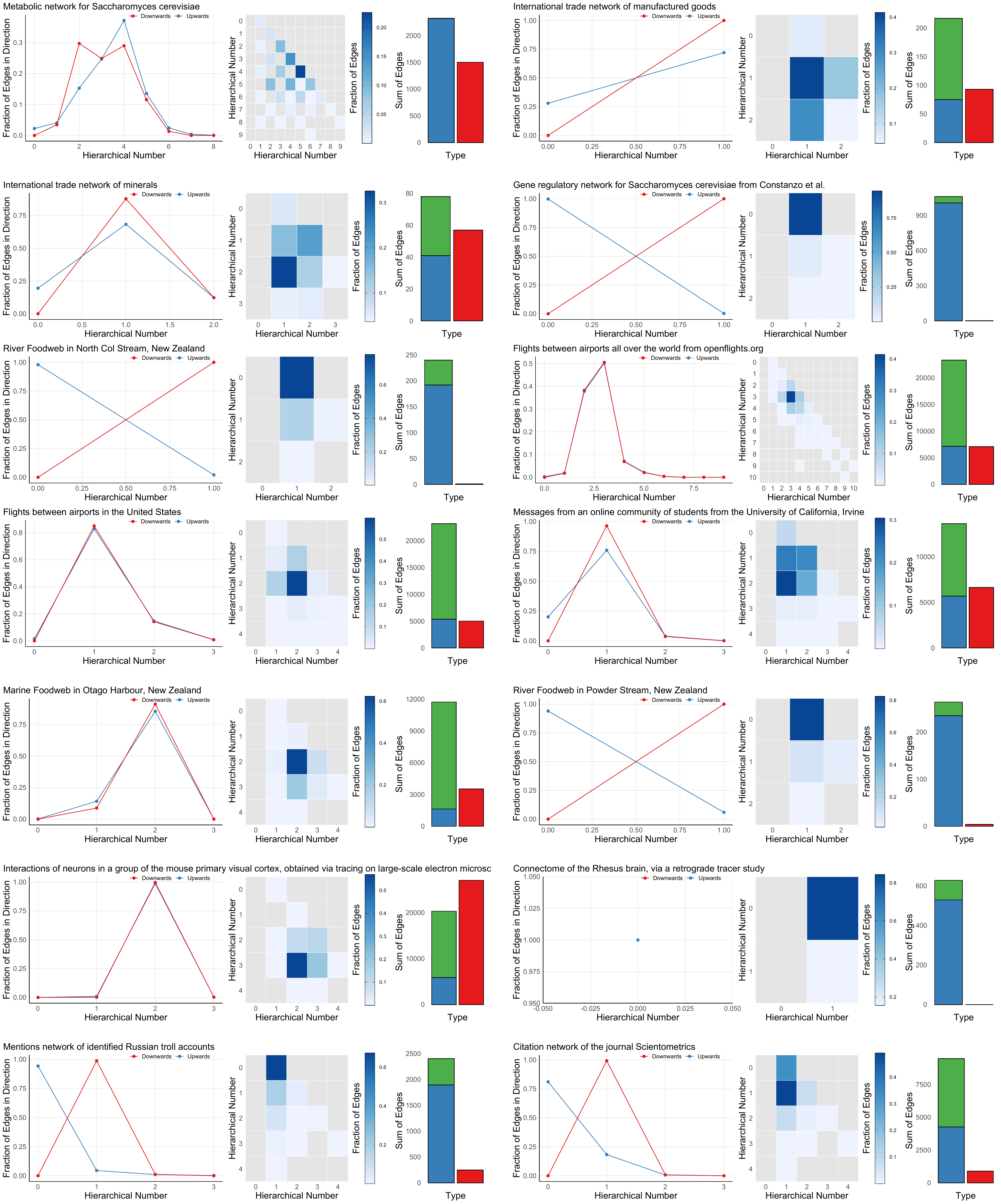}
	\end{figure}
	
	\begin{figure}
		\centering
		\includegraphics[width=\textwidth]{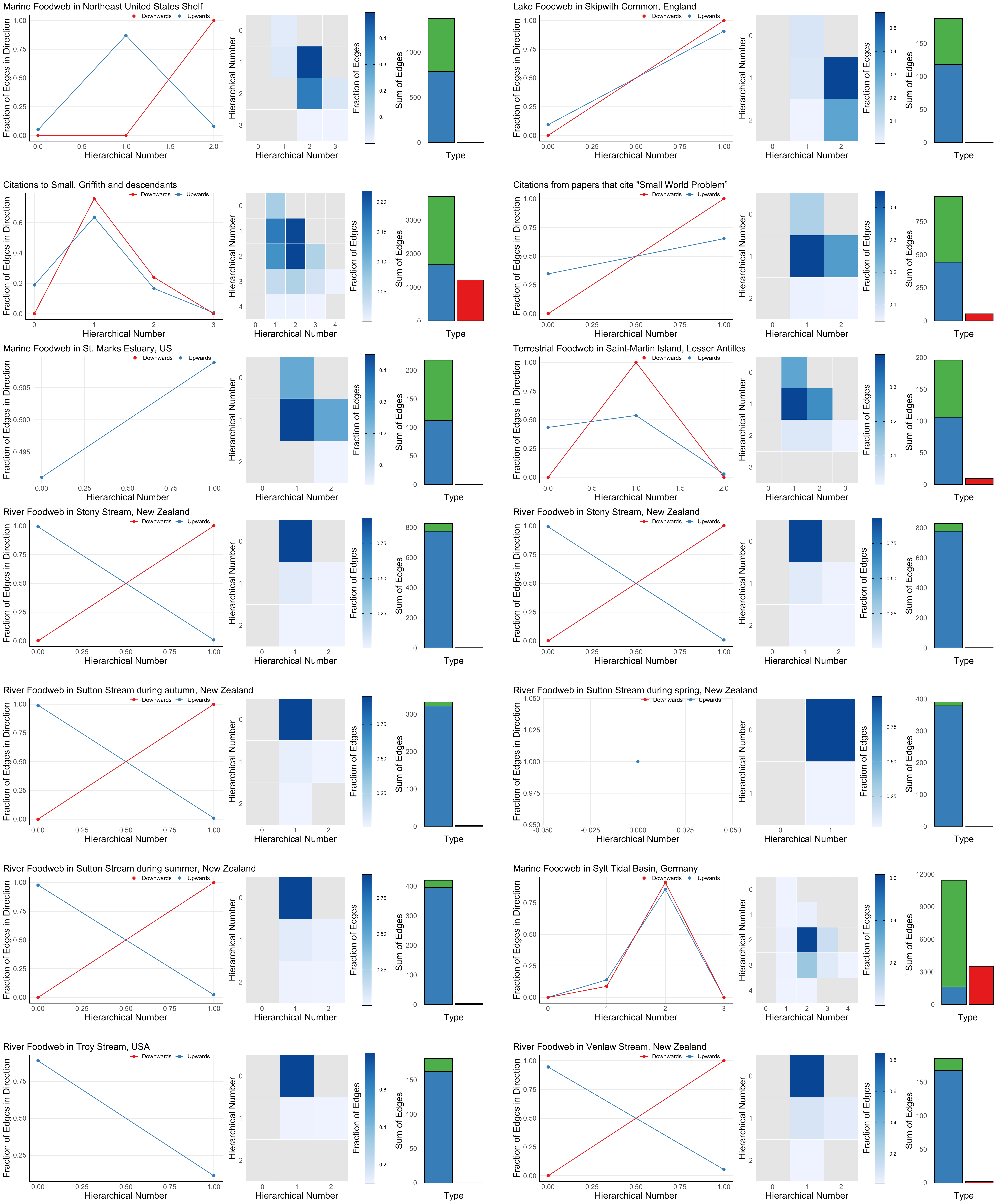}
	\end{figure}
	
	\begin{figure}
		\centering
		\includegraphics[width=\textwidth]{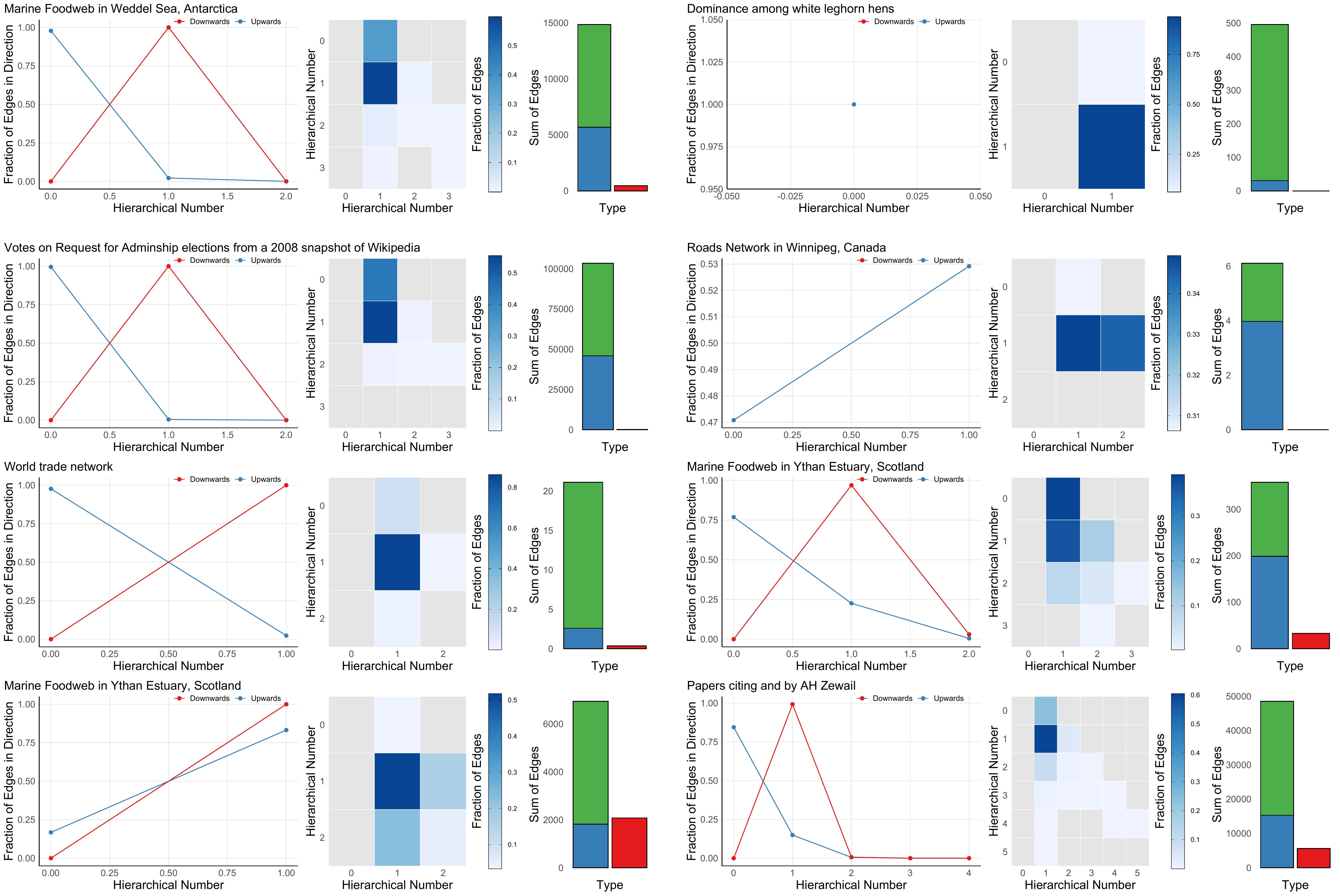}
	\end{figure}
	
\end{document}